\documentclass{elsart}
\usepackage{amssymb}
 
 \newtheorem{ttt}{\bfseries{Theorem}}[section]
 \newtheorem{ddd}{\bfseries{Definition}}
 \newtheorem{ppp}{\bfseries{Proposition}}[section]
 \newtheorem{lelele}{\bfseries{Lemma}}[section]
\journal{}
\begin{document}
\begin{frontmatter}
\title{Singularly Perturbed Self-Adjoint Operators in Scales of Hilbert spaces}
\author[f,g,r]{S. Albeverio}
\ead{albeverio@uni-bonn.de}
\author[s]{S. Kuzhel\corauthref{cor}}
\corauth[cor]{Corresponding author.}
\ead{kuzhel@imath.kiev.ua}
\author[s]{L. Nizhnik}
\ead{nizhnik@imath.kiev.ua}
\address[f]{Institut f\"{u}r Angewandte Mathematik, Universit\"{a}t Bonn, Wegelerstr. 6, D-53115 Bonn (Germany)}
\address[g]{SFB 611, Bonn, BiBoS, Bielefeld-Bonn}
\address[r]{CERFIM, Locarno and USI (Switzerland)}
\address[s]{Institute of Mathematics of the National Academy of
Sciences of Ukraine, Tereshchenkovskaya 3, 01601 Kiev (Ukraine)}

\begin{abstract}
Finite rank perturbations of a semi-bounded self-adjoint operator
$A$ are studied in the scale of Hilbert spaces associated with $A$.
A concept of quasi-boundary value space is used to describe
self-adjoint operator realizations of regular and singular
perturbations of $A$ by the same formula. As an application the
one-dimensional Schr\"{o}dinger operator with generalized zero-range
potential is considered in the Sobolev space $W^p_2(\mathbb{R})$,
$p\in\mathbb{N}$.
\end{abstract}
\end{frontmatter}
 \section{Introduction}
Let $A$ be a semibounded self-adjoint operator acting in a separable
Hilbert space $\mathcal{H}$ with inner product $(\cdot, \cdot)$ and
let $\mathcal{D}(A)$, $\mathcal{R}(A)$, and $\ker{A}$ denote the
domain, the range, and the null-space of $A$, respectively. Without
loss of generality, we will assume that $A\geq{I}$. Let
\begin{equation}\label{e2}
\mathcal{H}_s\subset\mathcal{H}=\mathcal{H}_0\subset\mathfrak{H}_{-s},
\ \ \ \ \ s>0
\end{equation}
be the standard scale of Hilbert spaces associated with $A$
($A$-scale) \cite{AL}, \cite{Ber}. Here, a Hilbert space
$\mathcal{H}_{s}$ $(s\in\mathbb{R})$ is considered as the completion
of the set $\cap_{n\in\mathbb{N}}{\mathcal D}({A^n})$ with respect
to the norm
\begin{equation}\label{e1}
\|u\|_{s}=\|A^{s/2}u\|, \ \ \ \ \ \ \ \
{u}\in\cap_{n\in\mathbb{N}}{\mathcal D}({A^n}).
\end{equation}

By (\ref{e1}), the operator $A^{r/2}$ \ $(r\in\mathbb{R})$ can
continuously be extended to an isometric mapping $A^{r/2}$ of
${\mathcal H}_{s}$ onto ${\mathcal H}_{s-r}$ (we preserve the same
notation $A^{r/2}$ for this continuation). In a natural way
${\mathcal H}_{s}$ and ${\mathcal H}_{-s}$ are dual and the inner
product in $\mathcal{H}$ can be extended to a pairing
\begin{equation}\label{e3}
<u, \psi>=(A^{s/2}u, A^{-s/2}\psi), \ \ \ u\in{{\mathcal H}_{s}}, \
\ \psi\in{\mathcal H}_{-s}
\end{equation}
such that $|<u, \psi>|\leq\|u\|_{s}\|\psi\|_{-s}$.

The present paper is an extended and modified variant of \cite{AKN}
and its aim consists in the development of a unified approach to the
study of finite rank perturbations of a self-adjoint operator $A$ in
the scale of Hilbert spaces $\mathcal{H}_s$.

We recall that a self-adjoint operator $\widetilde{A}\not=A$ acting
in ${\mathcal H}$ is called {\it a finite rank perturbation} of $A$
if the difference $(\widetilde{A}-z{I})^{-1}-({A}-z{I})^{-1}$ is a
finite rank operator in $\mathcal{H}$ for at least one point
$z\in{\mathbb{C}}\setminus{\mathbb{R}}$ \cite{Kato}.

If $\widetilde{A}$ is a finite rank perturbation of $A$, then the
corresponding symmetric operator\footnote{the symbol
$A\upharpoonright_{X}$ means the restriction of $A$ onto the set
$X$.}
\begin{equation}\label{ee4}
A_{\mathrm{sym}}={A}\upharpoonright_{\mathcal{D}}=\widetilde{A}\upharpoonright_{\mathcal{D}},
\ \ \ \ \
\mathcal{D}=\{u\in\mathcal{D}(A)\cap\mathcal{D}(\widetilde{A}) \ |\
Au=\widetilde{A}u\}
\end{equation}
arises naturally. This operator has finite and equal deficiency
numbers.

It is important that the operator $A_{\mathrm{sym}}$ can be
recovered uniquely by its defect subspace
$N=\mathcal{H}\ominus{\mathcal R}(A_{\mathrm{sym}})$ and the initial
operator $A$. Namely,
\begin{equation}\label{bonn3}
A_{\mathrm{sym}}=A\upharpoonright_{\mathcal{D}(A_{\mathrm{sym}})}, \
\ \mathcal{D}(A_\mathrm{sym})=\{u\in\mathcal{D}(A) \ | \ (Au,
\eta)=0, \ \forall{\eta}\in{N}\}
\end{equation}
Moreover, the choice of an arbitrary finite dimensional subspace
${N}$ of $\mathcal{H}$ as a defect subspace allows one to determine
by (\ref{bonn3}) a closed symmetric operator $A_{\mathrm{sym}}$ with finite and equal
defect numbers. To underline this relation, {\it we will use
notation} $A_N$ {\it instead of} $A_{\mathrm{sym}}$. Obviously, any
self-adjoint extension $\widetilde{A}$ of $A_N$ is a finite rank
perturbation of $A$.

A finite rank perturbation $\widetilde{A}$ of $A$ is called {\it
regular} if $\mathcal{D}(A)=\mathcal{D}(\widetilde{A})$. Otherwise
(i.e, $\mathcal{D}(A)\not=\mathcal{D}(\widetilde{A})$), the operator
$\widetilde{A}$ is called {\it singular}.

It is convenient to divide the class of singular perturbations into
two subclasses. We will say that a singular perturbation
$\widetilde{A}$ is {\it purely singular} if the symmetric operator
$A_{\mathrm{sym}}=A_N$ defined by (\ref{ee4}) is densely defined
(i.e., $N\cap\mathcal{D}(A)=\{0\}$) and {\it mixed singular} if
$A_N$ is nondensely defined (i.e., $N\cap\mathcal{D}(A)\not=\{0\}$).

Important examples of finite rank perturbations of the
Schr\"{o}dinger operator are given by finitely many point
interactions \cite{AL}, \cite{AL1}. The consideration of point
interactions in $L_2(\mathbb{R}^d)$ leads to purely singular
perturbations and, in the case of Sobolev spaces
$W_2^{p}(\mathbb{R}^d)$, $p\in\mathbb{N}$, mixed singular
perturbations arise \cite{NI}, \cite{Nizh}. These applications can
be served  as a certain motivation of the abstract results carried
out in the paper.

It is well-known that finite rank regular perturbations of $A$ can
be described with the help of finite rank self-adjoint operators
(potentials) acting in $\mathcal{H}$. Typical examples of finite
rank singular perturbations are provided by the general expression
 \begin{equation}\label{ne3}
 \widetilde{A}=A +V, \ \ \ \ \ \ V=\sum_{i,j=1}^{n}{b}_{ij}<\cdot,\psi_j>\psi_i \ \
 (\mathcal{R}(V)\not\subset\mathcal{H}, \ \
 b_{ij}\in\mathbb{C}).
 \end{equation}
Since $\mathcal{R}(V)\not\subset\mathcal{H}$, the singular potential
$V$ is not an operator in $\mathcal{H}$ and it acts in the spaces of
$A$-scale. Such types of expressions appear in many areas of
mathematical physics (for an extensive list of references, see
\cite{AL}, \cite{AL1}).

 In the present paper, we will study  finite rank singular
 perturbations of $A$ in the spaces of $A$-scale (\ref{e2}). {\it The main
 attention will be focused} on the description of self-adjoint
 extensions $\widetilde{A}$ of $A_{\mathrm{sym}}$ in a form
 that is maximally adapted for the determination of $\widetilde{A}$
 with the help of additive singular perturbations (\ref{ne3}) and
 preserves physically meaningful relations to the parameters
 $b_{ij}$ of the singular potential
 $V=\sum_{i,j=1}^{n}{b}_{ij}<\cdot,\psi_j>\psi_i$.

In Section 2, such a problem is solved for the case of purely
singular perturbations. Precisely, since the corresponding symmetric
operator $A_{\mathrm{sym}}=A_N$ in (\ref{ee4}) is densely defined,
we can combine the Albeverio -- Kurasov approach \cite{AL1} with the
boundary value spaces technique  \cite{Gor}, \cite{KK}. The first of
them allows us to involve the parameters $b_{ij}$ of the singular
potential in the determination of the corresponding self-adjoint
operator realization of (\ref{ne3}), the second provides convenient
framework for the description of such operators. As a result, we get
a simple description of self-adjoint realizations of purely singular
perturbations (Theorem \ref{ttt12}) and, moreover, we present a
simple algorithm for solving an inverse problem, i.e., recovering
the purely singular potential $V$ in (\ref{ne3}) by the given
self-adjoint extension of $A_N$ defined in terms of boundary value
spaces.

Other approaches to the description of purely singular perturbations
were recently suggested by Arlinskii and Tsekanovski \cite{ARTC} and
Posilicano \cite{P2}, \cite{P3}.

The description of mixed singular perturbations of $A$ is more
complicated because the corresponding symmetric operator $A_N$ is
nondensely defined and, hence, the adjoint of $A_N$ does not exist.
To overcome this problem, a certain generalization of the concept of
BVS is required. The key point here is the replacement of the
adjoint operator $A_N^*$ by a suitable object. In \cite{Der},
\cite{Mal}, the operator $A_N$ and its `adjoint' are understood as
linear relations and the description of all self-adjoint {\it
relations} that are extensions of the graph of $A_N$ was obtained.
In \cite{KK}, a pair of maximal dissipative extensions of $A_N$ and
its adjoint (maximal accumulative extension) was used instead of
$A_{N}^*$. This allows one to describe self-adjoint extensions
directly as operators without using linear relations technique.

The approaches mentioned above are general and they can be applied
to an arbitrary nondensely defined symmetric operator. However, in
the case where $A_N$ is determined as the restriction of an initial
self-adjoint operator $A$, it is natural to use $A$ for the
description of extensions of $A_N$ (see \cite{CD}, \cite{CO},
\cite{KO}). In Section 3, developing the ideas proposed recently in
\cite{NI}, \cite{Nizh}, we use $A$ for the definition of a
quasi-adjoint operator of $A_N$. The concept of quasi-adjoint
operators allows one to generalize the definition of boundary value
spaces (BVS) to the case of nondensely defined operators $A_ N$ and
to preserve the simple formulas for the description of self-adjoint
extensions of $A_N$.

One of the characteristic features of quasi-BVS extension theory
that immediately  follows from the definition of a quasi-BVS
consists in the description of {\it essentially}\footnote{i.e.,
those extensions that turn out to be self-adjoint after closure}
self-adjoint extensions of $A_N$. It should be noted that this
property is very convenient for the description of self-adjoint
differential expressions with complicated boundary conditions.
Furthermore, it gives the possibility to describe finite rank
regular and mixed singular perturbations of $A$ in just the same way
as purely singular perturbations.

In Section 4, the results of quasi-BVS extension theory are applied
to the study of finite rank singular perturbations of $A$ in spaces
of $A$-scale (\ref{e2}). In recent years, such kind of problems
attracted a steady interest and they naturally arise in the theory
of supersingular perturbations \cite{DHS}, \cite{Kura} and in the
study of Schr\"{o}dinger operators with point interactions in
Sobolev spaces \cite{NI}, \cite{Nizh}.

\setcounter{equation}{0}
\section{The Case of Purely Singular Perturbations}
\subsection{Description.}

In what follows we assume that $A\geq{I}$ is a self-adjoint operator
in $\mathcal{H}$, $N$ is a finite dimensional subspace of
$\mathcal{H}$, and $A_N$ is a symmetric operator defined by the
formula
\begin{equation}\label{e10}
A_{N}=A\upharpoonright_{\mathcal{D}(A_N)}, \ \ \ \
\mathcal{D}(A_N)=\{u\in\mathcal{D}(A) \ | \ (Au, \eta)=0, \
\forall{\eta}\in{N}\}.
\end{equation}

The operator $A_N$ is densely defined in ${\mathcal H}$ if and only
if $N\cap{\mathcal D}(A)=\{0\}$. In this case, ${\mathcal
D}(A_N^*)={\mathcal{D}}(A)\dot{+}N$ and
\begin{equation}\label{ee5}
A_N^*f=A_N^*(u+\eta)=Au, \ \ \ \forall{f}=u+\eta\in{\mathcal
D}(A_N^*) \ (u\in\mathcal{D}(A), \ \eta\in{N}).
\end{equation}

If $A_N$ is densely defined, then self-adjoint extensions of $A_N$
admit a convenient description in terms of boundary value spaces
(see \cite{GorKoch} and references therein).

\begin{ddd}\label{d1}
A triple $(\mathfrak{N}, \Gamma_0, \Gamma_1)$, where $\mathfrak{N}$
is an auxiliary Hilbert space and $\Gamma_0$, $\Gamma_1$ are linear
mappings of $\mathcal{D}(A_{\mathrm{sym}}^*)$ into $\mathfrak{N}$,
is called a boundary value space (BVS) of $A_N$ if the abstract
Green identity
\begin{equation}\label{tat15}
(A_N^*f, g)-(f, A_N^*g)=(\Gamma_1f,
\Gamma_0g)_{\mathfrak{N}}-(\Gamma_0f, \Gamma_1g)_{\mathfrak{N}}, \ \ \
\ f, g\in\mathcal{D}(A_N^*)
\end{equation}
is satisfied and the map $(\Gamma_0,
\Gamma_1):\mathcal{D}(A_N^*)\to\mathfrak{N}\oplus\mathfrak{N}$ is
surjective.
\end{ddd}

One of the simplest examples of BVS gives the triple\footnote{in
fact, this BVS was already implicitly used in the classical works
\cite{B1}, \cite{KR}} $(N, \Gamma_0, \Gamma_1)$, where ${N}$ is
taken from (\ref{e10}), (\ref{ee5}) and
\begin{equation}\label{tat234}
\Gamma_0(u+\eta)=P_{N}Au, \ \ \ \ \Gamma_1(u+\eta)=-\eta \ \ \ \
(\forall{u}\in{\mathcal{D}(A)}, \ \forall{\eta}\in{N}),
\end{equation}
where $P_N$ is the orthoprojector onto $N$ in ${\mathcal{H}}$.

The following elementary result enables one to get infinitely many
BVS of $A_N$ starting from the fixed one.
\begin{lelele}\label{ll32}
Let $(\mathfrak{N}, \Gamma_0, \Gamma_1)$ be a BVS of $A_N$ and let
$R$ be an arbitrary self-adjoint operator acting in $\mathfrak{N}$.
Then the triple $(\mathfrak{N}, \Gamma_0^{R}, \Gamma_1)$, where
$\Gamma_0^{R}=\Gamma_0-R\Gamma_1$ is also a BVS of $A_N$.
\end{lelele}

The next theorem provides a description of all self-adjoint
extensions of $A_N$.

\begin{ttt}[\cite{Koch}]\label{t1}
Let $(\mathfrak{N}, \Gamma_0, \Gamma_1)$ be a BVS of $A_N$. Then any
self-adjoint extension $\widetilde{A}$ of $A_N$ coincides with
restriction of $A_N^*$ to
\begin{equation}\label{k41}
{\mathcal{D}(\widetilde{A})}=\{f\in\mathcal{D}(A_N^*) \ | \
(I-U)\Gamma_0f=i(I+U)\Gamma_1f\},
\end{equation}
where $U$ is a unitary operator in $\mathfrak{N}$. Moreover, the
correspondence $\widetilde{A}\leftrightarrow{U}$ is a bijection
between the sets of all self-adjoint extensions of $A_N$ and all
unitary operators in $\mathfrak{N}$.
\end{ttt}

In cases where self-adjoint extensions are described by sufficiently
complicated boundary conditions (see, e.g., \cite{KS},), the
representation (\ref{k41}) is not always convenient because it
contains the same factor $U$ on the both sides. To overcome this
inconvenience, we outline another approach that enables one to
remove one of the factors in (\ref{k41}) but, simultaneously, to
preserve the description of all self-adjoint extensions of $A_N$.
The main idea here consists in the use of a family BVS
$(\mathfrak{N}, \Gamma_0^{R}, \Gamma_1)$ instead of a fixed BVS (see
\cite{KN} for details).

Let $(\mathfrak{N}, \Gamma_0^{R}, \Gamma_1)$ be a family of BVS of
$A_N$ defined in Lemma \ref{ll32}. For a fixed $R$, Theorem \ref{t1}
implies that the expression
\begin{equation}\label{tata77}
{A}_{B,R}:=A_{N}^{*}\upharpoonright_{\mathcal{D}({A}_{B,R})}, \ \ \
\mathcal{D}({A}_{B,R})=\{f\in\mathcal{D}(A_N^*) \ | \
{B}{\Gamma}_0^{R}f={\Gamma}_1f\},
\end{equation}
where $B$ is an arbitrary self-adjoint operator in $\mathfrak{N}$,
determines a subset $\mathcal{P}_R(A_N)$ of the set
$\mathcal{P}(A_N)$ of all self-adjoint extensions of $A_N$. More
precisely, a self-adjoint extension $\widetilde{A}$ of $A_N$ belongs
to $\mathcal{P}_R(A_N)\ \iff \
\mathcal{D}(\widetilde{A})\cap\ker{\Gamma}_0^{R}=\mathcal{D}(A_N).$

It is easy to verify, that the union $\bigcup_{R}\mathcal{P}_R(A_N)$
over all self-adjoint operators $R$ in $\mathfrak{N}$ coincides with
$\mathcal{P}(A_N)$. Moreover, for a fixed
$\widetilde{A}\in\mathcal{P}(A_N)$, there exist infinitely many $R$
such that $\widetilde{A}\in\mathcal{P}_R(A_N)$. Thus formula
(\ref{tata77}), where $R$ and ${B}$ play a role of parameters, gives
the description of all self-adjoint extensions of $A_N$.

\subsection{Self-adjoint realizations.}
\subsubsection{Construction of self-adjoint realizations by additive purely
singular perturbations.}
 Let us consider the general expression (\ref{ne3}),
 where $\psi_j$ $(1\leq{j}\leq{n})$ form a linearly
 independent system in ${\mathcal  H}_{-2}$ and
 the linear span $\mathcal{X}$ of $\{\psi_j\}_{j=1}^n$
 satisfies the condition
 $\mathcal{X}\cap{\mathcal H}=\{0\}$ (i.e., elements $\psi_j$
 are ${\mathcal H}$-independent).

 Let $\{e_j\}_1^n$ be the canonical basis of $\mathbb{C}^n$
 (i.e., $e_j=(0,\ldots,1,\ldots{0})$, where $1$ occurs on the $j$th place
 only). Putting $\Psi{e_j}:=\psi_j$ ($j=1,\ldots,n$), we define
 an injective linear mapping $\Psi:\mathbb{C}^n\to\mathcal{H}_{-2}$
 such that $\mathcal{R}(\Psi)=\mathcal{X}$.

 Let $\Psi^*:\mathcal{H}_2\to\mathbb{C}^n$  be the
 adjoint operator of $\Psi$ (in the sense \linebreak[4]
 $<u, \Psi{d}>=(\Psi^*u, d)_{\mathbb{C}^n}$,
 $\forall{u}\in\mathcal{H}_2, \forall{d}\in\mathbb{C}^n$).
 It is easy to see that
 \begin{equation}\label{nene}
 \Psi^*{u}=\left(\begin{array}{c}
 <u, \psi_1> \\
 \vdots \\
<u, \psi_n>
\end{array}\right), \ \ \ \ \forall{u}\in\mathcal{H}_{2}.
 \end{equation}
 Using (\ref{nene}), we rewrite the singular potential
 $V=\sum_{i,j=1}^{n}{b}_{ij}<\cdot,\psi_j>\psi_i$ in  (\ref{ne3}) as follows:
 \begin{equation}\label{ne33}
\sum_{i,j=1}^{n}{b}_{ij}<\cdot,\psi_j>\psi_i=\Psi\mathbf{B}\Psi^*,
 \end{equation}
 where the matrix $\mathbf{B}=\|b_{ij}\|_{i,j=1}^n$ consists of the
 coefficients $b_{ij}$ of the potential $V$. In what
 follows we assume that $\mathbf{B}$ is {\it Hermitian}, i.e.,
 $b_{ij}=\overline{b_{ji}}$.

 In order to give a meaning to $\widetilde{A}=A+V$ as a self-adjoint
 operator in ${\mathcal H}$  we consider a symmetric restriction
 $A_{\mathrm{sym}}$ of $A$
  \begin{equation}\label{e7}
 A_{\mathrm{sym}}:={A}\upharpoonright_{{\mathcal{D}}(A_{\mathrm{sym}})},   \ \ \ \
 {\mathcal{D}}(A_{\mathrm{sym}})={{\mathcal{D}}(A)}\cap\ker\Psi^*.
   \end{equation}

 By virtue of (\ref{e3}) (for $s=2$) and (\ref{nene}),
 the operator  $A_{\mathrm{sym}}$ is also defined
 by (\ref{e10}), where $A_{\mathrm{sym}}=A_N$ and
 $N=A^{-1}\mathcal{R}(\Psi)=A^{-1}\mathcal{X}$, i.e., $N$
 is a linear span of $\{A^{-1}\psi_j\}_{j=1}^n$.
 Since  $N\cap{\mathcal D}(A)=\{0\}$, the operator $A_N$ is densely
 defined in $\mathcal{H}$.

 Any self-adjoint extension $\widetilde{A}$ of $A_N$ is a purely
 singular perturbation of $A$ and, in general, it can be regarded as
 a realization of (\ref{ne3}) in $\mathcal{H}$.
 In this context, there arises the natural question of whether and how one
 could establish a physically meaningful correspondence between
 the parameter $\mathbf{B}$ of the potential
 $V=\Psi\mathbf{B}\Psi^*$ and self-adjoint extensions of $A_N$.

 To do this we combine the Albeverio--Kurasov approach
 \cite{AL1} with the BVS technique.  This approach
 consists in the construction of some regularization
\begin{equation}\label{tat44}
{A}_\mathrm{reg}:=A^++\Psi\mathbf{B}\Psi_{\mathbf{R}}^*=A^++\sum_{i,j=1}^n{b}_{ij}<\cdot,\psi_j^{\mathrm{ex}}>\psi_i,
\end{equation}
of (\ref{ne3}) that is well defined as
an operator from $\mathcal{D}(A_N^*)$ to
${\mathcal H}_{-2}$. (Here, $A^+$, $\Psi_{\mathbf{R}}^*$,
and $<\cdot,\psi_j^{\mathrm{ex}}>$
are extensions of $A$, $\Psi^*$, and $<\cdot,\psi_j>$ onto $\mathcal{D}(A_N^*)$).
After that, the corresponding self-adjoint realization $\widetilde{A}$
 of (\ref{ne3}) is determined by the formula
\begin{equation}\label{lesia40}
\widetilde{A}={A}_\mathrm{reg}\upharpoonright_{\mathcal{D}(\widetilde{A})}, \ \ \ \
\mathcal{D}(\widetilde{A})=\{f\in\mathcal{D}(A_{\mathrm{sym}}^*) \ | \
{A}_\mathrm{reg}f\in\mathcal{H}\}.
\end{equation}

By (\ref{ee5}), it is easy to see that for the definition of $A^+$
in (\ref{tat44}) one needs to determine the action of $A^+$ on $N$.
Assuming that $A^{+}\upharpoonright_{N}$ acts as the isometric
mapping $A$ in the $A$-scale, we get
\begin{equation}\label{tat101}
A^+f=Au+A\eta=A_N^*f+A\eta, \ \ \ \forall{f}=u+\eta\in\mathcal{D}(A_N^*).
\end{equation}

However, the principal point in the definition of ${A}_\mathrm{reg}$
is the construction of $\Psi_{\mathbf{R}}^*$ or, equivalently, the
definition of the functionals $<\cdot, \psi_j>$ $(j=1,\ldots,n)$ on
$\mathcal{D}(A_N^*)$.

It is clear (see (\ref{ee5})) that
 $<\cdot,\psi_j>$ can be extended onto $\mathcal{D}(A_N^*)$
 if we know its values on $N$.

 Since $N=A^{-1}\mathcal{R}(\Psi)$ and $\mathcal{R}(\Psi)$ coincides with
 the linear span of $\psi_j$ $(j=1, \ldots, n)$, the vectors
 $\eta_j={A}^{-1}\psi_j, \  j=1,\ldots,n$ form a basis of $N$. Using this fact and (\ref{ee5}),
 we get that any $f\in\mathcal{D}(A_N^*)$
can be represented as $f=u+\sum_{k=1}^n\alpha_k\eta_k$ \
$(u\in{\mathcal{D}}(A), \alpha_k\in\mathbb{C})$. Thus the extended
functional $<\cdot, \psi_j^{\mathrm{ex}}>$ is well-defined by the
formula
 \begin{equation}\label{k23}
<f, \psi_j^{\mathrm{ex}}>=<u,\psi_j>+\sum_{k=1}^n\alpha_k{r_{jk}}, \
\ \ \forall{f}\in\mathcal{D}(A_N^*)
 \end{equation}
if we know the entries $r_{jk}=<{A}^{-1}\psi_k, \psi_j,>=<\eta_k,
\psi_j,>$ of the regularization matrix
$\mathbf{R}=\|r_{jk}\|_{j,k=1}^n$. In this case, by virtue of
(\ref{nene}) and (\ref{k23}),
  \begin{equation}\label{ne1ne}
 \Psi^*_{\mathbf{R}}{f}=\Psi^*_{\mathbf{R}}(u+\sum_{k=1}^n\alpha_k\eta_k)
 =\Psi^*u+\mathbf{R}\left(\begin{array}{c}
 \alpha_1 \\
 \vdots \\
\alpha_n
\end{array}\right)=\left(\begin{array}{c}
 <f, \psi_1^{\mathrm{ex}}> \\
 \vdots \\
<f, \psi_n^{\mathrm{ex}}>
\end{array}\right)
 \end{equation}
for any ${f}\in\mathcal{D}(A_N^*)$.

If $\mathcal{R}(\Psi)\subset\mathcal{H}_{-1}$, the entries $r_{jk}$
are uniquely defined and $\mathbf{R}$ is an Hermitian matrix. In the
case where $\mathcal{R}(\Psi)\not\subset\mathcal{H}_{-1}$ the matrix
$\mathbf{R}$ is not determined uniquely \cite{AL1}.

In what follows we assume that $\mathbf{R}$ {\it is chosen as an
Hermitian matrix}.
\begin{lelele}\label{l23}
 The triple $({\mathbb C}^n, \Gamma_0^{\mathbf{R}}, \Gamma_1)$,
 where the linear operators $\Gamma_i^{\mathbf{R}}:\mathcal{D}(A_N^*)\to{\mathbb C}^n$
 are defined by the formulas
\begin{equation}\label{k9}
\Gamma_0^{\mathbf{R}}f=\Psi^*_{\mathbf{R}}f, \ \ \ \
\Gamma_1f=-\Psi^{-1}(A^+-A_N^*)f=-\Psi^{-1}A\eta
\end{equation}
(where $f=u+\eta, \ u\in\mathcal{D}(A), \eta\in{N}$) is a BVS of
$A_N$.
\end{lelele}

{\it{Proof.}} By (\ref{e3}), $<u, \psi_j>=(Au, \eta_j)$.
Taking into account this relation and
(\ref{ee5}), (\ref{nene}), (\ref{tat101})
 it is easy to verify that the mappings
 \begin{equation}\label{lesia99}
{\Gamma}_0f=\Psi^*u, \ \ \ \ \ \ {\Gamma}_1f=-\Psi^{-1}A\eta
\end{equation}
satisfy the conditions of Definition \ref{d1}.
Hence, $({\mathbb C}^n, {\Gamma}_0,
{\Gamma}_1)$ is a BVS of $A_N$.

It follows from (\ref{k23}), (\ref{nene}), (\ref{ne1ne}),
(\ref{k9}), and (\ref{lesia99}) that
$\Gamma_0^{\mathbf{R}}f={\Gamma}_0f-{\mathbf R}{\Gamma}_1f$. By
Lemma \ref{ll32} this means that $({\mathbb C}^n,
\Gamma_0^{\mathbf{R}}, \Gamma_1)$ is also a BVS of $A_N$. Lemma
\ref{l23} is proved.

 \begin{ttt}\label{ttt12}
Let $\widetilde{A}$ be a self-adjoint realization of (\ref{ne3})
defined by (\ref{tat44}), (\ref{lesia40}). Then
\begin{equation}\label{k4141}
\widetilde{A}=A_{\mathbf{B,
R}}=A_N^*\upharpoonright_{{\mathcal{D}(A_{\mathbf{B, R}})}}, \quad
{\mathcal{D}(A_{\mathbf{B, R}})}=\{f\in\mathcal{D}(A_N^*) \ | \
{\mathbf B}\Gamma_0^{\mathbf{R}}f=\Gamma_1f\},
\end{equation}
 $\Gamma_0^{\mathbf{R}}$ and $\Gamma_1$ being defined by (\ref{k9}).
\end{ttt}

{\it{Proof.}}  Employing relations
(\ref{tat44}), (\ref{tat101}), and  (\ref{k9}), we get
 $$
{A}_\mathrm{reg}f=A_N^*f+\Psi[{\mathbf
B}\Gamma_0^{\mathbf{R}}f-\Gamma_1f].
 $$
This equality and (\ref{lesia40}) mean that
$f\in\mathcal{D}(\widetilde{A})$ if and only if  ${\mathbf
B}\Gamma_0^{\mathbf{R}}f-\Gamma_1f=0$. Thus, the operator
realization $\widetilde{A}$ of (\ref{ne3}) coincides with the
operator $A_{\mathbf{B, R}}$ defined by (\ref{tata77}). Since
$\mathbf{B}$ is an Hermitian matrix, the operator $A_{\mathbf{B,
R}}$ is self-adjoint. Theorem \ref{ttt12}.

Summing the results above we can state that the choice of an
extension $\Psi^*_{\mathrm{ex}}$ of $\Psi^*$ onto
$\mathcal{D}(A_N^*)$ plays a main role and precisely this enables
one to choose (see (\ref{k9})) a more suitable\footnote{from the
point of view of the simplest relations between coefficients of
singular potentials and parameters of BVS.} BVS  $(\mathbb{C}^n,
\Gamma_0^{\mathbf R}, \Gamma_1)$ for the description of self-adjoint
realizations of (\ref{ne3}).

\subsubsection{Recovering purely singular
potentials by a given self-adjoint extension.} Here we consider an
inverse problem.  Namely, for a given BVS $(\mathbb{C}^n, \Gamma_0,
\Gamma_1)$ of $A_N$ such that $\ker\Gamma_1=\mathcal{D}(A)$ and the
corresponding self-adjoint extensions
\begin{equation}\label{a5}
{A}_{\mathbf{B}}={A}_N^*\upharpoonright_{\mathcal{D}({A}_{\mathbf{B}})},
\ \ \ \ {\mathcal{D}(A_{\mathbf{B}})}=\{f\in\mathcal{D}(A_N^*) \ | \
{\mathbf B}\Gamma_0f=\Gamma_1f\},
\end{equation}
where $\mathbf{B}$ is an Hermitian matrix, we recover an additive
purely singular perturbation $V=\Psi\mathbf{B}\Psi^*$ such that the
formal expression $\widetilde{A}=A+V$ possesses the self-adjoint
realization $A_{\mathbf{B}}$.

We start with the definition of $\Psi$. Since
$\ker\Gamma_1=\mathcal{D}(A)$, the restriction
$\Gamma_1\upharpoonright_{N}$  determines a one-to-one
correspondence between $N$ and $\mathbb{C}^n$. Hence,
$(\Gamma_1\upharpoonright_{N})^{-1}$ exists and
$(\Gamma_1\upharpoonright_{N})^{-1}$ maps $\mathbb{C}^n$ onto $N$.

Putting (cf. (\ref{k9}))
$\Psi{d}:=-A(\Gamma_1\upharpoonright_{N})^{-1}d$, where
$d\in\mathbb{C}^{n}$, we determine an injective linear mapping of
$\mathbb{C}^n$ to $\mathcal{H}_{-2}$ such that
$\mathcal{R}(\Psi)\cap\mathcal{H}=\{0\}$.

Set $\psi_j=\Psi{e_j}$, where $\{e_j\}_1^n$ is the canonical basis
of $\mathbb{C}^n$. Putting $f=u\in\mathcal{D}(A)$,
$g=A^{-1}\psi_j=A^{-1}\Psi{e_j}=-(\Gamma_1\upharpoonright_{N})^{-1}e_j$
in (\ref{tat15}) and recalling the condition
$\ker\Gamma_1=\mathcal{D}(A)$, we establish that $$ <u, \psi_j>=(Au,
A^{-1}\psi_j)=-(\Gamma_0u,
\Gamma_1A^{-1}\psi_j)_{\mathbb{C}^n}=(\Gamma_0u,
e_j)_{\mathbb{C}^n}. $$ This formula enables one to determine an
extension of $<\cdot, \psi_j>$ onto $\mathcal{D}(A_N^*)$ with the
help of the boundary operator $\Gamma_0$. Namely, $<f,
\psi_j^{\mathrm{ex}}>:=(\Gamma_0f, e_j)_{\mathbb{C}^n}$. But then,
reasoning by analogy with (\ref{ne1ne}), we conclude that
$\Gamma_0f=\Psi^*_{\mathbf{R}}f$.

Now, repeating arguments of Theorem \ref{ttt12}, it is easy to see
that the operator $A_{\mathbf{B}}$ defined by (\ref{a5}) is a
self-adjoint realization of the formal expression
$A^++\Psi\mathbf{B}\Psi_{\mathbf{R}}^*$.

{\bf Example 1.} {\it General zero-range potential in $\mathbb{R}$.}

 A one-dimensional Schr\"{o}dinger operator corresponding to a
 general zero-range potential at the point $x=0$ can be given by
 the formal expression
 \begin{equation}\label{ne2ne}
 -\frac{d^2}{dx^2}+b_{11}<\cdot,\delta>\delta+b_{12}<\cdot,\delta'>\delta+
 b_{21}<\cdot,\delta>\delta'+b_{22}<\cdot,\delta'>\delta',
  \end{equation}
 where $\delta'$ is the derivative of the Dirac
 $\delta$-function (with support at $0$) and the coefficients
 $b_{ij}$ form an Hermitian matrix.

Putting
$\Psi\left(\begin{array}{c}
1 \\
0
\end{array}\right)=\delta$ and $\Psi\left(\begin{array}{c}
0 \\
1
\end{array}\right)=\delta'$, we get
$\Psi^*u=\left(\begin{array}{c}
<u, \delta> \\
<u, \delta'>
\end{array}\right)$ ($u(x)\in{W}^2_2(\mathbb{R})$) and,
hence,
$$
\Psi{\mathbf{B}}\Psi^*=b_{11}<\cdot,\delta>\delta+b_{12}<\cdot,\delta'>\delta+
 b_{21}<\cdot,\delta>\delta'+b_{22}<\cdot,\delta'>\delta'
$$

In the case under consideration, $A=-d^2/dx^2+I$,
$\mathcal{D}(A)=W_2^2(\mathbb{R})$, where $W_2^2(\mathbb{R})$ is the
Sobolev space;
$A_{\mathrm{sym}}=(-d^2/dx^2+I)\upharpoonright_{\{u(x)\in{W}_2^2(\mathbb{R})
\mid u(0)=u'(0)=0\}}$ and $A_{\mathrm{sym}}=A_N$, where a subspace
$N$ of $L_2(\mathbb{R})$ is the linear span of functions
\begin{equation}\label{typ1}
\eta_1(x)={A}^{-1}\delta=\frac{1}{2}e^{-|x|}, \ \ \ \ \ \
\eta_2(x)={A}^{-1}\delta'(x)=-\frac{\mathrm{sign} \ {x}}{2}e^{-|x|}.
\end{equation}
Further $A_N^*f(x)=-f''(x)+f(x)$ \
($f(x)\in\mathcal{D}(A_N^*)={W_2^2}(\mathbb{R})\dot{+}N={W_2^2}(\mathbb{R}\backslash\{0\})$),
where the symbol $f''(x)$ means the second derivative (pointwise) of
$f(x)$ except the point $x=0$.

It follows from the description of $\mathcal{D}(A_N^*)$ that any
function $f\in\mathcal{D}(A_N^*)$ and its derivative $f'$ have
right(left)-side limits at the point $0$. Thus, the expressions
\begin{equation}\label{tat66}
g_r=\frac{g(+0)+g(-0)}{2}, \ \ \  g_s=g(+0)-g(-0),  \ \ (g=f \
\mbox{or} \ g=f')
\end{equation}
are well-posed. To obtain a regularization of (\ref{ne2ne}) it
suffices to extend the distributions $\delta$ and $\delta'$ onto
$\mathcal{D}(A_N^*)$. The most physically reasonable way, based on
the extension of $\delta$ by the continuity and parity onto
${W_2^2}(\mathbb{R}\backslash\{0\}$ and preserving the initial
homogeneity of $\delta'$ with respect to scaling transformations
\cite{AL1}, leads to the following extensions:\footnote{we omit
index $\mathrm{ex}$ for such natural extensions.}
$$ <f, \delta>=f_r, \ \ \ \ \ \ \ <f, \delta'>=-f'_r \ \ \ \ \ \
(f(x)\in{W_2^2}(\mathbb{R}\backslash\{0\}). $$

These extensions can also be determined by the general formula
(\ref{k23}), if we set $\mathbf{R}=\left(\begin{array}{cc}
 1/2 &  0 \\
 0 & -1/2
\end{array}\right)$.
In this case, $\Psi^*_{\mathbf{R}}f=\left(\begin{array}{c}
f_r \\
-f'_r
\end{array}\right)$ and the corresponding boundary operators
$\Gamma_0^{\mathbf{R}}$ and $\Gamma_1$ in the BVS $(\mathbb{C}^2,
\Gamma_0^{\mathbf{R}}, \Gamma_1)$ determined by (\ref{k9}) have the
form
\begin{equation}\label{tat67}
\Gamma_0^{\mathbf{R}}f(x)=\left(\begin{array}{c}
f_r \\
-f'_r
\end{array}\right), \ \ \  \Gamma_1f(x)=\left(\begin{array}{c}
f'_s \\
f_s
\end{array}\right), \ \ \
\forall{f(x)\in{W_2^2}(\mathbb{R}\backslash\{0\})}.
\end{equation}
Here the operator $\Gamma_0^{\mathbf{R}}f$ turns out to be the mean
value of $f(x)$ and $-f'(x)$ at the origin and $\Gamma_1$
characterizes the jumps of $f(x)$ and its derivative at the origin.

 Taking into account the fact that the operator
$A^+=-\frac{d^2}{dx^2}+I$ acts on
$f(x)\in{{W_2^2}(\mathbb{R}\backslash\{0\})}$ by the rule
$A^+f(x)=-\frac{d^2}{dx^2}f(x)+f(x)$, where the action of
$-\frac{d^2}{dx^2}f(x)$ is understood in the distributional sense,
i.e.,
$$
-\frac{d^2}{dx^2}f(x)=-f''(x)-f_s'\delta(x)-f_s\delta'(x)
$$
and employing Theorem \ref{ttt12} we obtain a description of
self-adjoint realizations $A_{\mathbf{B},\mathbf{R}}$ of
(\ref{ne2ne}) that are defined by the rule
$A_{\mathbf{B},\mathbf{R}}f(x)=-f''(x)$,
$$
f(x)\in\mathcal{D}(A_{\mathbf{B},\mathbf{R}})=\left\{f(x)\in{W}_2^2(\mathbb{R}\backslash\{0\})
\ | \
 \left(\begin{array}{cc}
 b_{11} & b_{12} \\
 b_{21} & b_{22}
\end{array}\right)\left(\begin{array}{c}
f_r \\
-f'_r
\end{array}\right)=\left(\begin{array}{c}
f'_s \\
f_s
\end{array}\right)\right\}.
$$

{\bf Example 2.} {\it  Point interaction in $\mathbb{R}^3$.}

Let us consider the self-adjoint operator $A=-\Delta+\mu^2I$,
$\mathcal{D}(A)={W_2^{2}(\mathbb{R}^{3})}$ acting in
$L_2(\mathbb{R}^3)$ and its restriction
$A_{\mathrm{sym}}=-\Delta+\mu^2I\upharpoonright_{\{u(x)\in{W_2^{2}(\mathbb{R}^{3})}
\mid u(0)=0 \}}$. It is easy to see that $A_{\mathrm{sym}}=A_N$,
where $N$ is the linear span of $\frac{e^{-\mu|x|}}{\mid{x}\mid}$
($\mu>0$).  The triple $(\mathbb{C}, \Gamma_0, \Gamma_1)$, where
\begin{equation}\label{a7}
\Gamma_1f=\lim_{\mid{x}\mid\to{0}}\mid{x}\mid{f}(x), \ \ \ \
\Gamma_0f=\lim_{\mid{x}\mid\to{0}}\left(f(x)-(\Gamma_1f)\frac{e^{-\mu|x|}}{\mid{x}\mid}\right)
 \end{equation}
$({f}(x)\in\mathcal{D}(A_N^*)={W_2^{2}(\mathbb{R}^{3})}\dot{+}N)$
forms a BVS of $A_N$. Moreover $\ker\Gamma_1=\mathcal{D}(A)$.

It follows from (\ref{a5}) and (\ref{a7}) that the operators
$$
A_b(u(x)+bu(0)\frac{e^{-\mu|x|}}{\mid{x}\mid})=(-\Delta+\mu^2I){u(x)},  \ \ \  \forall{u(x)}\in{W_2^{2}(\mathbb{R}^{3})}
$$
are self-adjoint extensions of $A_N$. By virtue of the results of
subsection 2.2.2, the operators $A_b$ can be considered as
self-adjoint realizations of the heuristic expression
$-\Delta+\mu^2+b<\cdot, \delta^{\mathrm{ex}}>\delta(x)$, where
$-\Delta$ is understood in the distributional sense and the
extension $\delta^{\mathrm{ex}}(x)$ of $\delta(x)$ is determined in
terms of the boundary operators $\Gamma_i$ as follows: $<f,
\delta^{\mathrm{ex}}>=\Gamma_0f$ \
($f\in{W_2^{2}(\mathbb{R}^{3})}\dot{+}N$)

\setcounter{equation}{0}
\section{The Case of Mixed Singular Perturbations}
\subsection{The concept of quasi-BVS.}
In the case of mixed singular perturbations, the operator $A_N$
determined by (\ref{e10}) is non-densely defined and its adjoint
operator $A_N^*$ does not exist. Thus some modification of BVS is
required to describe all self-adjoint extensions of $A_N$.

Let us suppose that there exists a real number $m>1$ such that
$N\cap\mathcal{D}(A^m)=\{0\}$. Then, the direct sum
\begin{equation}\label{bonn1}
{\mathcal{L}}_{m}:={\mathcal{D}}(A^m)\dot{+}N
\end{equation}
is well defined and we can define on ${\mathcal L}_{m}$ a
quasi-adjoint operator $A_N^{(*)}$ by the rule
\begin{equation}\label{e22}
A_{N}^{(*)}f=A_{N}^{(*)}(u+\eta)=Au, \  \ \
\forall{f}=u+\eta\in\mathcal{L}_m \ \ (u\in{\mathcal{D}(A^m)}, \
{\eta}\in{N}).
\end{equation}

Formula (\ref{e22}) is an analog of (\ref{ee5}) for the adjoint
operator $A_N^*$ and we can use $A_{N}^{(*)}$ as an analog of the
adjoint one.

It is easy to see that, in general,  $A_{N}^{(*)}$ is not closable
and it turns out to be closable only if $A_N$ is densely defined.

The concept of quasi-adjoint operators allows one to modify
Definition \ref{d1} and to extend it to the case of nondensely
defined symmetric operators.

\begin{ddd}\label{d2}
 A triple $(\mathfrak{N}, \Gamma_0, \Gamma_1)$, where
 $\Gamma_i$ are linear mappings of ${\mathcal L}_{m}$ in
 an auxiliary Hilbert space  $\mathfrak{N}$, is called  a
 quasi-BVS of $A_N$ if the abstract Green identity
\begin{equation}\label{new1}
(A_N^{(*)}f, g)-(f, A_N^{(*)}g)=(\Gamma_1f,
\Gamma_0g)_{\mathfrak{N}}-(\Gamma_0f, \Gamma_1g)_{{\mathfrak{N}}}, \
\ \ \forall{f, g}\in{\mathcal L}_{m}
\end{equation}
is satisfied and the map $(\Gamma_0,
\Gamma_1):\mathcal{L}_m\to\mathfrak{N}\oplus\mathfrak{N}$ is
surjective.
\end{ddd}
\begin{ppp}[\cite{KN}]\label{p34}
The following assertions are true:

1. If $A_N$ is densely defined, then an arbitrary BVS
$(\mathfrak{N}, \Gamma_0, \Gamma_1)$ of $A_N$ also is a quasi-BVS of
$A_N$.

2. If $A_N$ is nondensely defined, then the triple $(N,
\Gamma_0^{R}, \Gamma_1)$, where
\begin{equation}\label{tat99}
\Gamma_0^{R}(u+\eta)=P_NAu+R\eta, \ \ \ \Gamma_1(u+\eta)=-\eta \ \
(u\in\mathcal{D}(A^m), \ \eta\in{N})
\end{equation}
 is a quasi-BVS of $A_N$  for any choice of self-adjoint operator $R$ in $N$.

3. Let $(\mathfrak{N}, \Gamma_0, \Gamma_1)$ be a quasi-BVS of $A_N$.
Then the symmetric operator
\begin{equation}\label{e23}
A_N'=A_N^{(*)}\upharpoonright_{\mathcal{D}(A_N')}, \ \ \ \
\mathcal{D}(A_N')=\ker\Gamma_0\cap\ker\Gamma_1
\end{equation}
does not depend on the choice of quasi-BVS and its closure coincides
with $A_N$.
\end{ppp}

Let  $(\mathfrak{N}, \Gamma_0, \Gamma_1)$ be a quasi-BVS of $A_N$.
An unitary operator $U$ acting in $\mathfrak{N}$ is called {\it
admissible} with respect to $(\mathfrak{N}, \Gamma_0, \Gamma_1)$ if
the equation
\begin{equation}\label{e45}
(I-U)\Gamma_0f=i(I+U)\Gamma_1f, \ \ \ \
\forall{f}\in\mathcal{D}(A_N)\cap{\mathcal{L}_m}
\end{equation}
has only the trivial solution $\Gamma_0f=\Gamma_1f=0$.

If $A_N$ is densely defined, then
$\mathcal{D}(A_N)\cap{\mathcal{L}_m}=\mathcal{D}(A_N)\cap\mathcal{D}{(A^m)}=\mathcal{D}(A_N')$
and, by virtue of (\ref{e23}), any unitary operator $U$ in
$\mathfrak{N}$ is admissible. Otherwise ($A_N$ is nondensely
defined),
\begin{equation}\label{e47}
\mathcal{D}(A_N)\cap{\mathcal{L}_m}=\mathcal{D}(A_N')\dot{+}\mathcal{F},
\end{equation}
where $\dim{\mathcal{F}}=\dim({N\cap\mathcal{D}(A)})$.  Vectors
$f\in\mathcal{F}$ have the form $f=u+\eta$, where $\eta$ is an
arbitrary element of ${N\cap\mathcal{D}(A)}$ and $u$ is determined
by $\eta$ with the help of relation $P_N{A}(u+\eta)=0$ (this
determination is unique modulo $\mathcal{D}(A_N')$).

It follows from (\ref{e23}) and (\ref{e47}) that the condition of
admissibility takes away the lineal $\mathcal{F}$ from the set of
solutions of (\ref{e45}).

\begin{ttt}[cf. Theorem \ref{t1}]\label{t2}
Let $(\mathfrak{N}, \Gamma_0, \Gamma_1)$ be a quasi-BVS of $A_{N}$.
Then any self-adjoint extension $\widetilde{A}$ of $A_{N}$ is the
closure of the symmetric operator
\begin{equation}\label{k45}
\widetilde{A}'=A_{N}^{(*)}\upharpoonright_{\mathcal{D}(\widetilde{A}')},
\ {\mathcal{D}(\widetilde{A}')}=\{f\in\mathcal{D}(A_{N}^{(*)}) \ | \
(I-U)\Gamma_0f=i(I+U)\Gamma_1f\},
\end{equation}
where $U$ is an admissible unitary operator with respect to
$(\mathfrak{N}, \Gamma_0, \Gamma_1)$. Moreover, the correspondence
$\widetilde{A}\leftrightarrow{U}$ is a bijection between the set of
all self-adjoint extensions of $A_N$ and the set of all admissible
unitary operators.
\end{ttt}

{\it Proof.} Let $U$ be an admissible operator and let $\widetilde{A}'$
be the corresponding operator defined by (\ref{k45}). Since
$$
(\Gamma_1f, \Gamma_0g)_{\mathfrak{N}}-(\Gamma_0f,
\Gamma_1g)_{{\mathfrak{N}}}=
\frac{1}{2}\|(\Gamma_1+i\Gamma_0)f\|^2_{\mathfrak{N}}-
\frac{1}{2}\|(\Gamma_1-i\Gamma_0)g\|^2_{\mathfrak{N}},
$$
formula (\ref{new1}) implies that $\widetilde{A}'$ is a symmetric
extension of $A_N'$. Furthermore, there exists a linear subspace
${\mathcal{M}}$ of ${{\mathcal{L}}}_m$ such that
$\dim{{\mathcal{M}}}=\dim{\mathfrak{N}}=\dim{N}$ and
\begin{equation}\label{tat1}
{\mathcal{D}(\widetilde{A}')}={\mathcal{D}(A_N')}\dot{+}{\mathcal{M}}.
\end{equation}

It follows from the property of admissibility of $U$ and
(\ref{tat1}) that ${\mathcal{M}}\cap{{\mathcal{D}(A_N)}}={0}$. The
latter relation and assertion 3 of Proposition \ref{p34} mean that
$\widetilde{A}'$ is closable and its closure $\widetilde{A}$ is a
symmetric operator defined by the formula
\begin{equation}\label{k145}
\widetilde{A}=A_{N}^{(*)}\upharpoonright_{\mathcal{D}(\widetilde{A})},
\ \ \ \ \
{\mathcal{D}(\widetilde{A})}={\mathcal{D}(A_N)}\dot{+}{\mathcal{M}}.
\end{equation}

Since $\dim{{\mathcal{M}}}=\dim{N}$, the defect numbers of
$\widetilde{A}$ in the upper (lower) half plane are equal to $0$ and
hence, $\widetilde{A}$ is a self-adjoint extension of $A_N$. Thus we
show that the closure of $\widetilde{A}'$ defined by (\ref{k45}) is
a self-adjoint extension of $A_N$.

Conversely, let $\widetilde{A}$ be a self-adjoint extension of
$A_N$. It follows from Theorem 5.15 (\cite[Chapter 1]{KK}) that
$\widetilde{A}$ is determined by (\ref{k145}), where
${\mathcal{M}}\subset{{\mathcal{L}}}_m$ and
$\dim{{\mathcal{M}}}=\dim{N}$.
 But then the symmetric operator
 $\widetilde{A}'=\widetilde{A}\upharpoonright_{\mathcal{D}(\widetilde{A})\cap\mathcal{L}_m}$
 defined by (\ref{tat1}) is an essentially self-adjoint restriction
of $\widetilde{A}$. The domain
${\mathcal{D}(\widetilde{A}')}={\mathcal{D}(\widetilde{A})\cap\mathcal{L}_m}$
admits the representation (\ref{k45}), where the admissibility of
$U$ follows from the relation
${\mathcal{M}}\cap{{\mathcal{D}(A_N)}}={0}$ and the unitarity of $U$
follows from the property of $\widetilde{A}$ to be a self-adjoint
operator. Theorem \ref{t2} is proved.

{\bf Remark.} If $U$ is not admissible, then the domain
$\mathcal{D}(\widetilde{A}')$ of a symmetric operator
$\widetilde{A}'$ defined by (\ref{k45}) has a nontrivial intersection
with $\mathcal{F}$ and $\widetilde{A}'$ is not closable.

By analogy with the densely defined case we can describe
self-adjoint extensions of $A_N$ as the closure of the symmetric
operators
\begin{equation}\label{e6}
{A}_{B}'=A_{N}^{(*)}\upharpoonright_{\mathcal{D}({A}_{B}')}, \ \ \ \
\ {\mathcal{D}({A}_{B}')}=\{f\in\mathcal{L}_m \ | \
B\Gamma_0f=\Gamma_1f\},
\end{equation}
where $(\mathfrak{N}, \Gamma_0, \Gamma_1)$ is a quasi-BVS and $B$ is
a self-adjoint operator in $\mathfrak{N}$. In such a setting, the
operator $B$ is called {\it admissible} with respect to
$(\mathfrak{N}, \Gamma_0, \Gamma_1)$ if the equation
$B\Gamma_0f=\Gamma_1f$ ($f\in\mathcal{D}(A_N)\cap\mathcal{L}_m$) has
only the trivial solution $\Gamma_0f=\Gamma_1f=0$.
\begin{ppp}[\cite{KN}]\label{p42}
If $B$ is an admissible operator, then the closure of ${A}_{B}'$ is
a self-adjoint extension of $A_N$.

A self-adjoint extension $\widetilde{A}$ of $A_N$ can be represented
as the closure of a symmetric operator $A_{B}'$ defined by
(\ref{e6}) if and only if
$\mathcal{D}(\widetilde{A})\cap\ker\Gamma_0=\mathcal{D}(A_{N}')$.
\end{ppp}

Since (\ref{e6}) does not describe all self-adjoint extensions of
$A_N$, a situation where any operator $B$  is admissible in
(\ref{e6}) is possible.

\begin{ppp}\label{ppp21}
If $(\mathfrak{N}, {\Gamma}_0, {\Gamma}_1)$ is a quasi-BVS of $A_N$
such that $\ker{\Gamma}_0\supset\mathcal{D}(A_N)\cap\mathcal{L}_m$,
then the closure of ${A}_{B}'$ defined by (\ref{e6}) is a
self-adjoint extension of $A_N$ for any self-adjoint operator $B$ in
$\mathfrak{N}$.
\end{ppp}

{\it Proof.} If
$\ker{\Gamma}_0\supset\mathcal{D}(A_N)\cap\mathcal{L}_m$, then the
equation $B{\Gamma}_0f={\Gamma}_1f$
$(f\in\mathcal{D}(A_N)\cap\mathcal{L}_m)$ has only the trivial
solution ${\Gamma}_0f={\Gamma}_1f=0$ and hence, any self-adjoint
operator $B$ is admissible with respect to $(\mathfrak{N},
{\Gamma}_0, {\Gamma}_1)$. Proposition \ref{ppp21} is proved.

Let us specify the obtained results and present more constructive
condition of admissibility for the family of quasi-BVS $(N,
\Gamma_0^{R}, \Gamma_1)$ determined by (\ref{tat99}).
\begin{ppp}\label{p10}
 1. A self-adjoint operator $B$ acting in $N$ is admissible with
 respect to $(N, \Gamma_0^{R}, \Gamma_1)$ if and only if the equation
\begin{equation}\label{tat9}
BP_{N}A\eta=(I+BR)\eta, \ \ \  \forall{\eta}\in{N\cap\mathcal{D}(A)}
\end{equation}
has the unique solution $\eta=0$.

 2. Formula (\ref{e6}) (where $\Gamma_0=\Gamma_0^{R}$) determines
 self-adjoint extensions of $A_N$ for any choice of $B$
 if and only if the operator $R$ satisfies the relation
 $P_NA\eta=R\eta$ for all $\eta\in{N}\cap\mathcal{D}(A)$.
 \end{ppp}

 {\it Proof.} Assertion 1 follows directly from (\ref{tat99}) and
 the description of the elements of $\mathcal{F}\subset\mathcal{D}(A_N)\cap\mathcal{L}_m$.
 To establish assertion 2, it suffices to observe that
$\Gamma_0^{R}f=P_N{Au}+R\eta=-P_N{A}\eta+R\eta$
 for all elements $f=u+\eta\in\mathcal{F}$. Thus,
 $$
 \ker{\Gamma}_0^{R}\supset\mathcal{F}\iff
 P_NA\eta=R\eta \ \ \mbox{for all} \ \ \eta\in{N}\cap\mathcal{D}(A).
 $$
 Employing now Proposition \ref{ppp21}, we complete the proof.

{\bf Example 3.} Let us consider a Schr\"{o}dinger operator that is
determined by analogy with (\ref{ne2ne}), where $\delta'$ is
replaced by a function $q\in{L_2(\mathbb{R})}$:
 \begin{equation}\label{ne3ne}
-\frac{d^2}{dx^2}+b_{11}<\cdot,\delta>\delta+b_{12}(\cdot, q)\delta+
 b_{21}<\cdot,\delta>q+b_{22}(\cdot, q)q.
  \end{equation}

In our case, $A=-d^2/dx^2+I$, $\mathcal{D}(A)=W_2^2(\mathbb{R})$ and
the defect subspace $N\subset{L_2(\mathbb{R})}$ is the linear span
of the functions  $\eta_1(x)={A}^{-1}\delta=\frac{1}{2}e^{-|x|}, \ \
 \eta_2(x)={A}^{-1}q(x)$.

For the sake of simplicity, we assume that the function $q(x)$
coincides with a fundamental solution $\mathfrak{m}_{2k}(x)$
($k\geq1$) of the equation
$(-d^2/dx^2+I)^{k}{\mathfrak{m}_{2k}}(x)=\delta$. In this case,
$\eta_1=\mathfrak{m}_{2}$, $\eta_2=\mathfrak{m}_{2k+2}$.

Let us fix $m=k+1$, then, according to (\ref{bonn1}),
$\mathcal{L}_m={W}_2^{2k+2}(\mathbb{R})\dot{+}N\subset{{W_2^{2k+2}}(\mathbb{R}\backslash\{0\})}$.
It is easy to see that an arbitrary function $f\in\mathcal{L}_m$
admits the representation
$$
f(x)=u(x)-f_s'\mathfrak{m}_{2}(x)-f_s^{[2k+1]}\mathfrak{m}_{2k+2},
$$
where $u\in{W}_2^{2k+2}(\mathbb{R})$ and $f_s'$ and $f_s^{[2k+1]}$
mean the jumps of the functions $f'(x)$ and $f^{[2k+1]}(x)$ at the
point $x=0$. Here,
$f^{[2k+1]}(x):=\frac{d}{dx}(-\frac{d^2}{dx^2}+I)^{k}f(x)$ \
$(x\not=0)$.

By the direct verification, we get that the triple $(\mathbb{C}^2,
\Gamma_0, \Gamma_1)$, where
$$
\Gamma_0f(x)=\left(\begin{array}{c}
f(0) \\
(f, \mathfrak{m}_{2})
\end{array}\right), \ \ \  \Gamma_1f(x)=\left(\begin{array}{c}
f'_s \\
f_s^{[2k+1]}
\end{array}\right), \ \ \
\forall{f(x)}\in\mathcal{L}_m
$$
is a quasi-BVS of $A_N$.

In our case, all conditions of Proposition \ref{ppp21} are satisfied
and, hence, the restriction of $A_N^{(*)}$
($A_N^{(*)}f(x)=-f''(x)+f(x),$ $x\not=0$) onto the collection of
functions $f\in\mathcal{L}_m$ that are specified by the boundary
conditions
$$
f'_s=b_{11}f(0)+b_{12}(f, \mathfrak{m}_{2}), \ \ \
f_s^{[2k+1]}=b_{21}f(0)+b_{22}(f, \mathfrak{m}_{2})
$$
is an essentially self-adjoint operator in $L_2(\mathbb{R})$. The
closure of such an operator has the form $A_{q}+I$, where $A_{q}$ is
a self-adjoint realization of the heuristic expression (\ref{ne3ne})
The operator $A_{q}$ can be interpreted as the Schr\"{o}dinger
operator with nonlocal point interaction \cite{AN5}. Its domain
$\mathcal{D}(A_q)$ consists of all functions
$f\in{{W_2^{2}}(\mathbb{R}\backslash\{0\})}$ that satisfy the
boundary conditions  $f_s=0,$ \ $f'_s=b_{11}f(0)+b_{12}(f, q)$ and
the action of $A_{q}f$ is determined as follows:
$$
A_{q}f=-f''(x)+b_{21}q(x)f(0)+b_{22}(f, q)q(x), \ \ \ x\not=0.
$$

\subsection{Quasi-BVS and finite rank regular perturbations.} Here
we are going to show that the concept of quasi-BVS enables one to
describe finite rank regular perturbations of $A$ in just the same
way as finite rank purely singular perturbations.  To illustrate
this point, we consider the following one-dimensional regular
perturbation:
\begin{equation}\label{ee33}
A_\alpha=A+\alpha(\cdot, \psi)\psi,\ \ \  \psi\in{\mathcal{H}}_s\setminus{\mathcal{H}}_{s+\epsilon}\ (\forall\epsilon>0).
\end{equation}
The rank one operator $\alpha(\cdot, \psi)\psi$ is a bounded
operator in $\mathcal{H}$ and the operator $A_\alpha$ is
self-adjoint on the domain ${\mathcal{D}(A)}$.

On the other hand, we can consider $A_\alpha$ and $A$ as
two self-adjoint extensions of the symmetric nondensely defined operator
(cf. (\ref{e10}))
\begin{equation}\label{tat8}
A_N=A\upharpoonright_{\mathcal{D}(A_N)}, \ \
\mathcal{D}(A_N)=\{u\in\mathcal{D}(A) \ | \ (u,\psi)=(Au, A^{-1}\psi)=0\}.
\end{equation}
Here $N$ is the linear span of $\eta=A^{-1}\psi$ (i.e., $N=<\eta>$)
and
$\eta\in{\mathcal{H}}_{s+2}\setminus{\mathcal{H}}_{s+2+\epsilon}$.

Let us describe self-adjoint extensions of $A_N$. To do this, we
fix $m>s+2$ and consider the direct sum
$\mathcal{L}_m={\mathcal{D}}(A^m)\dot{+}<\eta>.$

In what follows, without loss of generality we assume that
$\|\eta\|=1$. Then, any element $f\in\mathcal{L}_m$ admits the
presentation $f=u+\beta\eta$, where ${u}\in{\mathcal{D}}(A^m)$ and
$\beta\in\mathbb{C}$ and the operators $\Gamma_0^{R}, \Gamma_1$
defined by (\ref{tat99}) have the form\footnote{we use the notation
$r$ instead of $R$ to emphasize that $R$ is an operator
multiplication by a real number $r$.}
 $$
\Gamma_0^{R}(u+\beta\eta)=P_NAu+r\beta\eta=[(Au,\eta)+r\beta]\eta, \
\ \ \ \Gamma_1(u+\beta\eta)=-\beta\eta,
$$
where the parameter $r$ is an arbitrary real number.

The triple $(N, \Gamma_0^{R}, \Gamma_1)$ is a quasi-BVS of $A_N$ and
Theorem \ref{t2} gives the description of all self-adjoint
extensions of $A_N$. In particular, formula (\ref{e6}) (where
$\Gamma_0=\Gamma_0^{R}$) shows that the closure of operators
\begin{equation}\label{tat10}
A_b'f=A_b'(u+\beta\eta)=Au, \ \ \
\mathcal{D}(A_b')=\{f=u+\beta\eta \ \ | \ \
b[(Au,\eta)+r\beta]=-\beta\}
\end{equation}
are self-adjoint extensions of $A_N$ and they coincide with
operators $A_\alpha$ (see (\ref{ee33})) if we put
$$
b=\frac{\alpha}{1+\alpha[(A\eta,\eta)-r]}.
$$
In particular, if $r=(A\eta,\eta)$, then $b=\alpha$.

\setcounter{equation}{0}
\section{Finite Rank Singular Perturbations of $A$ in spaces of $A$-Scale}

Let $p$ be a fixed integer ($p\in\mathbb{N})$. Since
$\mathcal{H}_{p}$ is a Hilbert space, all known results on finite
rank perturbations of $A$ can automatically be reformulated for its
image $A\upharpoonright_{\mathcal{D}(A^{p/2+1})}$ acting in
${\mathcal H}_{p}$ as a self-adjoint operator. However, the specific
of ${\mathcal H}_{p}$ as a space of the $A$-scale (\ref{e2}) enables
one to get a lot of new nontrivial results (see, e.g., \cite{NI}
\cite{Nizh}, where the spectral analysis of Schr\"{o}dinger
operators with point interactions in the Sobolev spaces
$W^{p}_2(\mathbb{R}^d)$ was carried out). The aim of this section is
to generalize the results of \cite{NI}, \cite{Nizh} for the abstract
case of a self-adjoint operator acting in ${\mathcal{H}}_{p}$.

\subsection{Construction of BVS for powers of $A_N$.}
Let $N$ be a finite dimensional subspace of ${\mathcal H}$ such that
$N\cap{\mathcal{D}}(A)=\{0\}$ and let $A_N$ be the corresponding
symmetric densely defined operator constructed by $N$ (see
(\ref{e10})).

The following statement shows that an arbitrary power of $A_N$ is a
symmetric restriction of the same power of $A$ defined by the
special choice of a defect subspace $\widetilde{M}$ in
$\mathcal{H}$.
\begin{lelele}\label{l3}
For any $p\in\mathbb{N}$,  $A_N^{p+1}:=(A_N)^{p+1}$ is a symmetric
densely defined operator in $\mathcal{H}$ and
$A_N^{p+1}=(A^{p+1})_{\widetilde{M}}$, where
$\widetilde{M}=N\dot{+}A^{-1}N\dot{+}\ldots,\dot{+}A^{-p}N$ and
$$
(A^{p+1})_{\widetilde{M}}=A^{p+1}\upharpoonright_{\mathcal{D}((A^{p+1})_{\widetilde{M}})},
\ \mathcal{D}((A^{p+1})_{\widetilde{M}})={\{u\in\mathcal{D}(A^{p+1})
| (A^{p+1}u, \mathfrak{m})=0,
\forall{\mathfrak{m}}\in{\widetilde{M}}\}}.
$$
\end{lelele}

{\it Proof.} Since $\mathcal{D}(A^{p+1})\cap{\widetilde{M}}=\{0\}$,
the operator $(A^{p+1})_{\widetilde{M}}$ is densely defined. To
prove $A_N^{p+1}=(A^{p+1})_{\widetilde{M}}$ it suffices to observe
that
$\mathcal{D}(A_N^{p+1})=\mathcal{D}((A^{p+1})_{\widetilde{M}})$.
Lemma \ref{l3} is proved.

The next statement gives a convenient algorithm for the construction
of BVS of $A_N^{p+1}$ starting from a fixed BVS of $A_N$.

\begin{ttt}\label{ttt44}
Let $(\mathfrak{N}, \Gamma_0, \Gamma_1)$ be a BVS of $A_N$ and let
$p\in\mathbb{N}$. Then the triple $(\oplus\mathfrak{N}^{p+1},
\widetilde{\Gamma}_0, \widetilde{\Gamma}_1)$, where
$\oplus\mathfrak{N}^{p+1}:=\underbrace{\mathfrak{N}\oplus\mathfrak{N}\oplus\ldots\oplus\mathfrak{N}}_{p+1
\ \mbox{times}}$ and
\begin{equation}\label{tat14}
\widetilde{\Gamma}_0{f}=\left(\begin{array}{c}
\Gamma_0{f} \\
\Gamma_0A_N^*{f} \\
\vdots  \\
\Gamma_0(A_N^*)^{p}{f}
\end{array}\right), \ \ \ \
\widetilde{\Gamma}_1{f}=\left(\begin{array}{c}
\Gamma_1(A_N^*)^{p}{f} \\
\Gamma_1(A_N^*)^{p-1}{f} \\
\vdots  \\
\Gamma_1{f}
\end{array}\right), \ \ \ \forall{{f}}\in\mathcal{D}((A_N^*)^{p+1})
\end{equation}
is a BVS of $A_N^{p+1}$.
\end{ttt}

{\it Proof.} It follows from Lemma \ref{l3} that
$((A_N)^{p+1})^*=(A_N^*)^{p+1}$. Hence, the operators
$\widetilde{\Gamma}_i$ are well defined on
$\mathcal{D}(((A_N)^{p+1})^*)=\mathcal{D}((A_N^*)^{p+1})$.
Furthermore employing (\ref{tat15}) and (\ref{tat14}) we directly
verify the following equality for any
${f},{g}\in\mathcal{D}((A_N^*)^{p+1})$:
$$
\begin{array}{l}
((A_N^*)^{p+1}{f}, {g})-({f}, (A_N^*)^{p+1}{g})=((A_N^*)^{p+1}{f},
{g})-((A_N^*)^{p}{f}, A_N^*{g})+ \\ ((A_N^*)^{p}{f},
A_N^*{g})-((A_N^*)^{p-1}{f}, (A_N^*)^2{g})+\ldots +(A_N^*{f},
(A_N^*)^{p}{g})-({f}, (A_N^*)^{p+1}{g})=\\
(\Gamma_1(A_N^*)^{p}{f}, \Gamma_0{g})_{\mathfrak{N}}-
(\Gamma_0(A_N^*)^{p}{f}, \Gamma_1{g})_{\mathfrak{N}}+
(\Gamma_1(A_N^*)^{p-1}{f}, \Gamma_0(A_N^*)^{2}{g})_{\mathfrak{N}}-
\\ (\Gamma_0(A_N^*)^{p-1}{f},
\Gamma_1(A_N^*)^{2}{g})_{\mathfrak{N}}+\ldots (\Gamma_1{f},
\Gamma_0(A_N^*)^{p}{g})_{\mathfrak{N}}- (\Gamma_0{f},
\Gamma_1(A_N^*)^{p}{g})_{\mathfrak{N}}= \\ (\widetilde{\Gamma}_1{f},
\widetilde{\Gamma}_0{g})_{\oplus\mathfrak{N}^{p+1}}
-(\widetilde{\Gamma}_0{f},
\widetilde{\Gamma}_1{g})_{\oplus\mathfrak{N}^{p+1}}
\end{array}.
$$

To prove that  $(\widetilde{\Gamma}_0, \widetilde{\Gamma}_1)$ maps
$\mathcal{D}((A_N^*)^{p+1})$ onto
$(\oplus\mathfrak{N}^{p+1})\oplus(\oplus\mathfrak{N}^{p+1})$  some
auxiliary preparations are required.

At first, the property of $(\mathfrak{N}, \Gamma_0, \Gamma_1)$ to
be a BVS of $A_N$ and (\ref{ee5}) yield
\begin{equation}\label{ttato1}
\mathcal{D}(A_N^{p+1})=\ker\widetilde{\Gamma}_0\cap\ker\widetilde{\Gamma}_1.
\end{equation}

Further, since $\mathcal{D}(A_N^p)$ is dense in $\mathcal{H}$ and
$\dim{N}<\infty$, the relation $P_N\mathcal{D}(A_N^p)=N$ ($P_N$ is
the orthoprojector onto $N$ in $\mathcal{H}$) holds for any
$p\in{\mathbb{N}}$. This equality enables one to verify (with the
use of (\ref{e10})) that
$A^{-1}\mathcal{D}(A_N^{p})+\mathcal{D}(A_N)\supset{A^{-1}N}$. But
then recalling that $\mathcal{D}(A)=\mathcal{D}(A_N)\dot{+}A^{-1}N$
we get
\begin{equation}\label{ttato2}
A^{-1}\mathcal{D}(A_N^{p})+\mathcal{D}(A_N)+N=\mathcal{D}(A)\dot{+}N
=\mathcal{D}(A_N^*).
\end{equation}

Let us prove the surjective property of the map
$(\widetilde{\Gamma}_0, \widetilde{\Gamma}_1)$ for $p=1$. To do this
we present an arbitrary vectors  $\widetilde{F}_0,
\widetilde{F}_1\in\oplus\mathfrak{N}^{2}=\mathfrak{N}\oplus\mathfrak{N}$
as the vector columns $\widetilde{F}_i=(F_{i0},
F_{i1})^{\mathrm{t}}$ ($i=0,1$ and $\mathrm{t}$ denotes the
transposition). Then equations
$\widetilde{\Gamma}_if=\widetilde{F}_i$
($f\in\mathcal{D}((A_N^*)^{2})$) are equivalent to the following
system of equations:
\begin{equation}\label{ttato3}
{\Gamma}_if=F_{i0}, \ \ \ \ \ {\Gamma}_iA_N^*f=F_{i1}, \ \ \
 f\in\mathcal{D}((A_N^*)^{2}) \ \ \ i=0,1.
\end{equation}

Since $(\mathfrak{N}, \Gamma_0, \Gamma_1)$ is a BVS of $A_N$, there
exists $g'\in\mathcal{D}(A_N^*)$ such that
\begin{equation}\label{ttato4}
{\Gamma}_ig'=F_{i0}, \ \ \ \ i=0, 1.
\end{equation}
It is important that such $g'$ is not defined uniquely. Precisely,
by virtue of (\ref{ttato1}), any $g=g'+u$, where
$u\in\mathcal{D}(A_N)$ satisfy (\ref{ttato4}).

Let us consider the element $f=A^{-1}g+\eta=A^{-1}g'+A^{-1}u+\eta$,
where $u\in\mathcal{D}(A_N)$ and $\eta\in{N}$ are arbitrary
elements. Clearly, $f\in\mathcal{D}((A_N^*)^2)$ and, by
(\ref{ttato4}), ${\Gamma}_iA_N^*f=F_{i1}$ $(i=0,1)$.

Taking into account the definition of $f$, we can rewrite the rest
equations of (\ref{ttato3}) as follows:
$$
\Gamma_0(A^{-1}u+\eta)=F_{00}-\Gamma_0A^{-1}g', \ \ \ \ \
\Gamma_1(A^{-1}u+\eta)=F_{10}-\Gamma_1A^{-1}g',
$$
where $u\in\mathcal{D}(A_N)$ and $\eta\in{N}$ play the role of
`free' variables. Employing now (\ref{ttato2}) for $p=1$ and
recalling the equality
$\mathcal{D}(A_N)=\ker{\Gamma}_0\cap\ker{\Gamma}_1$ we conclude that
the latter two equations have a solution for a certain choice of
vectors $u=u_s$ and $\eta=\eta_s$. So, we prove that
$f=A^{-1}g'+A^{-1}u_s+\eta_s$ is a solution of (\ref{ttato3}).
Hence, $(\widetilde{F}_0, \widetilde{F}_1)$ maps
$\mathcal{D}((A_N^*)^{2})$ onto
$(\oplus\mathfrak{N}^{2})\oplus(\oplus\mathfrak{N}^{2})$.

The general case $p\in\mathbb{N}$ is verified by the induction.
Theorem \ref{ttt44} is proved.

{\bf Example 4}. Let $\mathcal{H}=L_2(\mathbb{R})$, $A=-d^2/dx^2+I$,
$\mathcal{D}(A)=W_2^2(\mathbb{R})$ and let $A_N$ and
$({\mathbb{C}}^2, \Gamma_0^R, \Gamma_1)$ be a symmetric operator and
its BVS, respectively, that are defined in Example 1 (see
(\ref{tat67})). In this case,
$\mathcal{D}(A_N^*)={W}_2^{2}(\mathbb{R}\setminus\{0\})$ and
$A_N^*f(x)=-d^2f(x)/dx^2+f(x)$
($f(x)\in{W}_2^{2}(\mathbb{R}\setminus\{0\}),\ x\not=0 $).

Let $p\in\mathbb{N}$. Then  $A_N^{p+1}=(-d^2/dx^2+I)^{p+1}$,
$$
\mathcal{D}(A_N^{p+1})=
{\{u(x)\in{W}_2^{2p+2}(\mathbb{R}) \mid u(0)=u'(0)=\ldots=u^{(2p)}(0)=u^{(2p+1)}(0)=0\}}
$$
and $(A_N^*)^{p+1}f(x)=(-d^2/dx^2+I)^{p+1}f(x)$ ($x\not=0$) for all
$f(x)\in{W}_2^{2p+2}(\mathbb{R}\setminus\{0\})$.

To simplify the notation we will use the following symbol for
quasi-derivatives of
$f(x)\in{{W}_2^{2p+2}(\mathbb{R}\setminus\{0\})}$:
$$
f^{[2k]}(x):=\left(-\frac{d^2}{dx^2}+I\right)^{k}f(x), \ \ \
f^{[2k+1]}(x):=\frac{d}{dx}f^{[2k]}(x), \ \ k\in\mathbb{N}\cup{0}.
$$ Thus $(A_N^*)^{p+1}f(x)=f^{[2p+2]}(x)$.

According to Theorem \ref{ttt44} and (\ref{tat67}), a triple
$({\mathbb{C}}^{2p+2}, \widetilde{\Gamma}_0, \widetilde{\Gamma}_1)$,
where $$ \widetilde{\Gamma}_0{f}=\left(\begin{array}{c} f_{r} \\
-f_r^{[1]}
\\ \vdots  \\ {f_r}^{[2p]} \\ -f_r^{[2p+1]}
\end{array}\right), \ \ \ \ \ \
\widetilde{\Gamma}_1{f}=\left(\begin{array}{c}
{f_s}^{[2p+1]} \\
{f_s}^{[2p]} \\
\vdots  \\
f_s^{[1]} \\
f_s
\end{array}\right) \ \ \
(f(x)\in{W}_2^{2p+2}(\mathbb{R}\setminus\{0\}))
$$ is a BVS of $A_N^{p+1}$. Here the indexes $r$ and $s$ mean,
respectively, the mean value and the jump at $x=0$ of the
corresponding quasi-derivative $f^{[\tau]}(x)$ (see (\ref{tat66})).
The Green identity related to $({\mathbb{C}}^{2p+2},
\widetilde{\Gamma}_0, \widetilde{\Gamma}_1)$ has the form
$$
(f^{[2p+2]}, g)_{L_2(\mathbb{R})}-(f, g^{[2p+2]})_{L_2(\mathbb{R})}=
\sum_{\tau=0}^{2p+1}(-1)^{\tau}f_r^{[\tau]}\overline{g_s^{[2p+1-\tau]}}-
\sum_{\tau=0}^{2p+1}(-1)^{\tau}f_s^{[2p+1-\tau]}\overline{g_r^{[\tau]}},
$$ where $f$ and $g$ are arbitrary functions from
${{W}_2^{2p+2}(\mathbb{R}\setminus\{0\})}$ \cite{Nizh}.

\subsection{Construction of quasi-BVS for a symmetric operator $A_M$
in ${\mathcal H}_{p}$.}

As was noted above, the self-adjoint operator
$A_p:=A\upharpoonright_{\mathcal{D}(A^{p/2+1})}$ acting in
$\mathcal{H}_p$ can be considered as an image of the initial
operator $A_0:=A$ in $\mathcal{H}_p$. In this case
$\mathcal{D}(A_p)=\mathcal{D}(A^{p/2+1})$. By analogy with
(\ref{e10}), we fix a finite dimensional subspace $M$ of ${\mathcal
H}_{p}$ and determine a symmetric operator
\begin{equation}\label{eee10}
A_M=A_p\upharpoonright_{\mathcal{D}(A_M)}, \ \
\mathcal{D}(A_M)=\{u\in\mathcal{D}(A_p) \ | \ (A_pu,
{\mathfrak{m}})_p=0, \  \forall\mathfrak{m}\in{M}\}
\end{equation}
acting in ${\mathcal H}_p$. In this subsection, we will consider the
case where
\begin{equation}\label{tat19}
M=\sum_{k=0}^{p/2}\dot{+}A^{-p+k}N:=A^{-p}N\dot{+}A^{-p+1}N\dot{+}\ldots\dot{+}A^{-\frac{p}{2}}N.
\end{equation}
Here $p$ is assumed to be \emph{even} and $N$ is a finite
dimensional subspace of $\mathcal{H}$ such that
$N\cap\mathcal{D}(A)=\{0\}$.

For such a choice of $M$ the definition (\ref{eee10}) of $A_M$ can
be rewritten as follows: $A_M=A\upharpoonright_{\mathcal{D}(A_M)},$
\begin{equation}\label{tyt20}
\mathcal{D}(A_M)=\{u\in\mathcal{D}(A^{p/2+1}) \ | \
P_NAu=P_NA^2u=\ldots=P_NA^{p/2+1}u=0\},
\end{equation}
 where $P_N$ is the
orthoprojector onto $N$ in $\mathcal{H}$ or, that is equivalent,
\begin{equation}\label{tat38}
A_M=A_N\upharpoonright_{\mathcal{D}(A_M)}, \ \ \ \ \
\mathcal{D}(A_M)=\mathcal{D}(A_N^{p/2+1}).
\end{equation}
Thus the operator $A_M$ is closely related to $A_N$ defined by
(\ref{e10}).

It follows from (\ref{tat19}) that
$M\cap\mathcal{D}(A^{p/2+1})\supset{A^{-p}N}\not=\{0\}$. Hence,
$A_M$ is a nondensely defined symmetric operator in ${\mathcal H}_p$
and for it we can construct a quasi-BVS only. To do this, we chose
$m=(p+1)/(p/2+1)$. Then $\mathcal{D}({A}_p^m)=\mathcal{D}(A^{p+1})$,
 the direct sum ${\mathcal
L}_m=\mathcal{D}({A}_p^m)\dot{+}M=\mathcal{D}(A^{p+1})\dot{+}M $ is
well posed and we can define the action of $A_M^{(*)}f$ on any
element $f=u+\mathfrak{m}\in\mathcal{L}_m$ by the formula (cf.
(\ref{e22}))
\begin{equation}\label{e2233}
A_M^{(*)}f=A_{M}^{(*)}(u+\mathfrak{m})=A_pu=Au, \qquad
\forall{u}\in\mathcal{D}(A^{p+1}), \quad
\forall{\mathfrak{m}}\in{M}.
\end{equation}
\begin{ttt}\label{t37}
Let $A_N$ be defined by (\ref{e10}) and let $(\mathfrak{N},
\Gamma_0, \Gamma_1)$ be a BVS of $A_N$ such that
$\ker\Gamma_1=\mathcal{D}(A)$. Then the triple
$(\oplus\mathfrak{N}^{p/2+1}, \widehat{\Gamma}_0,
\widehat{\Gamma}_1)$, where
$\oplus\mathfrak{N}^{p/2+1}=\underbrace{\mathfrak{N}\oplus\mathfrak{N}\oplus\ldots\oplus\mathfrak{N}}_{p/2+1
\ \mbox{times}}$ and
\begin{equation}\label{tat17}
\widehat{\Gamma}_0{f}=\left(\begin{array}{c}
\Gamma_0f \\
\Gamma_0A_N^*{f} \\ \vdots  \\ \Gamma_0(A_N^*)^{\frac{p}{2}}{f}
\end{array}\right), \ \ \
\widehat{\Gamma}_1{f}=\left(\begin{array}{c}
\Gamma_1(A_N^*)^{p}{f} \\ \Gamma_1(A_N^*)^{p-1}{f} \\ \vdots  \\
\Gamma_1(A_N^*)^{\frac{p}{2}}{f}
\end{array}\right), \ \  \forall{{f}}\in\mathcal{L}_m
\end{equation}
is a quasi-BVS of the symmetric operator $A_M$ in $\mathcal{H}_p$.
In particular, the Green identity
\begin{equation}\label{tyt9}
(A_M^{(*)}{f}, {g})_p-({f}, A_M^{(*)}{g})_p= (\widehat{\Gamma}_1{f},
\widehat{\Gamma}_0{g})_{\oplus\mathfrak{N}^{p/2+1}}
-(\widehat{\Gamma}_0{f},
\widehat{\Gamma}_1{g})_{\oplus\mathfrak{N}^{p/2+1}}
\end{equation}
 is true for any $f, g\in\mathcal{L}_m=\mathcal{D}(A^{p+1})\dot{+}M$.
\end{ttt}

{\it Proof.}  It follows from Lemma \ref{l3} that
$\mathcal{D}((A_N^*)^{p+1})=\mathcal{D}(A^{p+1})\dot{+}\widetilde{M}$
and $(A_N^*)^{p+1}(u+\widetilde{\mathfrak{m}})=A^{p+1}u$, where
$u\in\mathcal{D}(A^{p+1})$ and
$\widetilde{\mathfrak{m}}\in\widetilde{M}$. By virtue of
(\ref{tat19}), $M=\widetilde{M}\cap\mathcal{H}_p$. Thus, the latter
relations
 and (\ref{e2233}) imply that
\begin{equation}\label{tat21}
A_M^{(*)}{f}=A^{-p}(A_N^*)^{p+1}{f}
\end{equation}
for any
${f}\in\mathcal{L}_m=\mathcal{D}(A_{M}^{(*)})=\mathcal{D}((A_N^*)^{p+1})\cap\mathcal{H}_p.$

Using the assumption that $\ker\Gamma_1=\mathcal{D}(A)$ and
relations (\ref{tat14}), (\ref{tat21}), we verify the abstract Green
identity for any $f, g\in\mathcal{L}_m$:
\begin{eqnarray*}
\lefteqn{((A_N^*)^{p+1}{f}, {g})-({f}, (A_N^*)^{p+1}{g})=
(A^{-p}(A_N^*)^{p+1}{f}, {g})_p-({f}, A^{-p}(A_N^*)^{p+1}{g})_p=}
\nonumber \\ \lefteqn{(A_M^{(*)}{f}, {g})_p-({f}, A_M^{(*)}{g})_p=
(\widehat{\Gamma}_1{f},
\widehat{\Gamma}_0{g})_{\oplus\mathfrak{N}^{p/2+1}}
-(\widehat{\Gamma}_0{f},
\widehat{\Gamma}_1{g})_{\oplus\mathfrak{N}^{p/2+1}}.}
\\ & & \mbox{}
\end{eqnarray*}

Let $F_0$, $F_1$ be an arbitrary elements from
$\oplus\mathfrak{N}^{p/2+1}$. Since $\oplus\mathfrak{N}^{p/2+1}$ can
be embedded into $\oplus\mathfrak{N}^{p+1}$ as a subspace
$({\oplus\mathfrak{N}^{p/2+1}})\oplus\underbrace{0\oplus\ldots,\oplus,
0}_{\frac{p}{2} \ \mathrm{times}}$, the elements $F_i$ belong to
$\oplus\mathfrak{N}^{p+1}$ and have the representations: $$
F_0=(\underbrace{\eta_1^0, \eta_2^0, \ldots, \eta_{p/2+1}^0, 0,
\ldots, 0}_{p+1 \ \mathrm{times}}), \ \ \ F_1=(\underbrace{\eta_1^1,
\eta_2^1, \ldots, \eta_{p/2+1}^1, 0, \ldots, 0}_{p+1 \
\mathrm{times}}). $$

Since $(\oplus\mathfrak{N}^{p}, \widetilde{\Gamma}_0,
\widetilde{\Gamma}_1)$ is a BVS of $A_N^{p+1}$ constructed in
Theorem \ref{ttt44}, there exists $f\in\mathcal{D}((A_N^*)^{p+1})$
such that $\widetilde{\Gamma}_0f=F_0$ and
$\widetilde{\Gamma}_1f=F_1$. Furthermore, it follows from
(\ref{tat14}) and the choice of $F_1$ that
$\Gamma_1f=\ldots=\Gamma_1(A_N^*)^{\frac{p}{2}-1}f=0$. These
equalities and condition $\ker\Gamma_1=\mathcal{D}(A)$ mean that
$f\in\mathcal{D}(A^{p/2})=\mathcal{H}_p$. But then, the description
of $\mathcal{L}_m$ in (\ref{tat21}) implies that
$f\in{\mathcal{L}_m}$. To complete the proof of Theorem \ref{t37} it
suffices to observe that
$\widetilde{\Gamma}_if=\widehat{\Gamma}_if$, where
$\widehat{\Gamma}_i$ have the form (\ref{tat17}).

{\bf Example 5.} (cf. Example 4). Let $\mathcal{H}=L_2(\mathbb{R})$,
$A=-d^2/dx^2+I$, $\mathcal{D}(A)=W_2^2(\mathbb{R})$ and let $A_N$ be
a symmetric operator defined in Example 1. In this case,
$\mathcal{H}_p$ coincides with the Sobolev space
$W^p_2(\mathbb{R})$, $p\in\mathbb{N}$. Further, by (\ref{tat38}),
the symmetric operator $A_M$ acting in $W^p_2(\mathbb{R})$ has the
form $A_M=-d^2/dx^2+I$, $$
\mathcal{D}(A_M)=\{u(x)\in{W}_2^{p+2}(\mathbb{R}) \ | \
u(0)=u'(0)=\ldots=u^{(p)}(0)=u^{(p+1)}(0)=0 \}. $$ Here, the defect
subspace $M$ is determined by (\ref{tat19}) and it coincides with a
linear span of fundamental solutions $\mathfrak{m}_{2j}(x)$ of the
equation $(-d^2/dx^2+I)^{j}{\mathfrak{m}_{2j}}(x)=\delta$ and their
derivatives $\mathfrak{m}_{2j-1}(x)=\mathfrak{m}'_{2j}(x)$ that
belong to $\mathcal{H}_p$. Precisely, $M$ is a linear span of the
functions $$ {\mathfrak{m}}_2j(x)=
\frac{1}{(j-1)!2^{j}}\sum_{r=0}^{j-1}C^r_{2j-2-r}(2j-3-2r)!!|x|^re^{-|x|},
 \ \ \mathfrak{m}_{2j-1}(x)=\mathfrak{m}'_{2j}(x),
$$
where index $j$ runs the set $\{p/2+1, p/2+2, \ldots, p+1\}$.

 The operator $A_M$ is nondensely defined in
$W^p_2(\mathbb{R})$. Its quasi-adjoint $A_M^{(*)}$ (see
(\ref{e2233}) and (\ref{tat21})) is defined on the domain $$
\mathcal{D}(A_M^{(*)})=\mathcal{L}_m={W^{2p+2}_2(\mathbb{R})}\dot{+}M=W^p_2(\mathbb{R})\cap{W^{2p+2}_2(\mathbb{R}\setminus\{0\})}
$$ and acts as follows: $A_M^{(*)}f(x)=A^{-p}f^{[2p+2]}(x)$ for
all
$f(x)\in{W}^p_2(\mathbb{R})\cap{W^{2p+2}_2(\mathbb{R}\setminus\{0\})}$.

Let $({\mathbb{C}}^2, \Gamma_0^{\mathbf{R}}, \Gamma_1)$ be a BVS of
$A_N$ defined by (\ref{tat67}). Obviously,
$\ker\Gamma_1=\mathcal{D}(A)$. According to Theorem \ref{t37}, the
triple $({\mathbb{C}}^{p+2}, \widehat{\Gamma}_0,
\widehat{\Gamma}_1)$, where
\begin{equation}\label{ene23}
\widehat{\Gamma}_0{f}=\left(\begin{array}{c} f_{r} \\
-f_r^{[1]}
\\ \vdots  \\
f^{[p]}_r  \\
-f^{[p+1]}_r
\end{array}\right), \ \ \
\widehat{\Gamma}_1{f}=\left(\begin{array}{c} {f_s}^{[2p+1]}
\\ {f_s}^{[2p]} \\ \vdots  \\
f_s^{[p+1]} \\
f_s^{[p]}
\end{array}\right)
\end{equation}
$(f(x)\in{W}^p_2(\mathbb{R})\cap{W^{2p+2}_2(\mathbb{R}\setminus\{0\})})$
is a quasi-BVS of the symmetric operator $A_M$ acting in
$W^p_2(\mathbb{R})$. The corresponding Green identity has the form
$$ (A_M^{(*)}f, g)_{W_2^p(\mathbb{R})}-(f,
A_M^{(*)}g)_{W_2^p(\mathbb{R})}=
\sum_{\tau=0}^{p+1}(-1)^{\tau}f_r^{[\tau]}\overline{g_s^{[2p+1-\tau]}}-
\sum_{\tau=0}^{p+1}(-1)^{\tau}f_s^{[2p+1-\tau]}\overline{g_r^{[\tau]}},
$$ where $f$ and $g$ are arbitrary functions from
$W^p_2(\mathbb{R})\cap{W^{2p+2}_2(\mathbb{R}\setminus\{0\})}$
(\cite{Nizh}).

\subsection{Description of self-adjoint extensions of $A_M$ in $\mathcal{H}_p$.}

A quasi-BVS $(\oplus\mathfrak{N}^{p/2+1}, \widehat{\Gamma}_0,
\widehat{\Gamma}_1)$ of $A_M$ presented in Theorem \ref{t37} enables
one to get a simple description of self-adjoint extensions of $A_M$
in $\mathcal{H}_p$.

\begin{lelele}\label{lela1}
Let $\widehat{\Gamma}_0$ be determined by (\ref{tat17}). Then
$\ker\widehat{\Gamma}_0\supset\mathcal{D}(A_M)\cap\mathcal{L}_m$.
\end{lelele}

{\it Proof.} Obviously $\ker{\Gamma}_0\supset\mathcal{D}(A_N)$
(since $(\mathfrak{N}, \Gamma_0, \Gamma_1)$ is a BVS of $A_N$). But
then relations (\ref{tat38}) and (\ref{tat17}) give that
$\widehat{\Gamma}_0f=0$ for any
$f\in\mathcal{D}(A_M)\cap\mathcal{L}_m$. Lemma \ref{lela1} is
proved.

By Lemma \ref{lela1}, the equation
$B\widehat{\Gamma}_0f=\widehat{\Gamma}_1f$ \
($f\in\mathcal{D}(A_M)\cap\mathcal{L}_m$) has only the trivial
solution $\widehat{\Gamma}_0f=\widehat{\Gamma}_1f=0$ for an
arbitrary self-adjoint operator $B$ acting in
$\oplus\mathfrak{N}^{p/2+1}$. So, any $B$ is admissible with respect
to the quasi-BVS $(\oplus\mathfrak{N}^{p/2+1}, \widehat{\Gamma}_0,
\widehat{\Gamma}_1)$.

The next statement is a direct consequence of Proposition
\ref{ppp21}.

\begin{ttt}\label{p25}
For an arbitrary self-adjoint operator ${B}$ in
$\oplus\mathfrak{N}^{p/2+1}$ the formula
\begin{equation}\label{tat77}
{A}_{B}'=A_{M}^{(*)}\upharpoonright_{\mathcal{D}({A}_{{B}}')}, \ \ \
\mathcal{D}({A}_{{B}}')=\{f\in\mathcal{L}_m \ | \
{B}\widehat{\Gamma}_0f= \widehat{\Gamma}_1f\},
\end{equation}
determines an essentially self-adjoint operator in $\mathcal{H}_{p}$
and its closure is a self-adjoint extension of $A_M$ in
$\mathcal{H}_{p}$.
\end{ttt}

{\bf Example 6.} Let us preserve the notation of Example 5 and let
$\mathbf{B}$ be an arbitrary Hermitian matrix of the order $p+2$.
 Then, according to Theorem \ref{p25}, the closure of the operator
 ${A}_{{\mathbf{B}}}'$ defined by the rule:
${A}_{{\mathbf{B}}}'f(x)=A^{-p}f^{[2p+2]}(x)$, where $f(x)$ belong
to ${W}^p_2(\mathbb{R})\cap{W^{2p+2}_2(\mathbb{R}\setminus\{0\})}$
and satisfy the condition
$$
\mathbf{B}\widehat{\Gamma}_0{f}=\widehat{\Gamma}_1{f} \ \ \ \ \ \ \
(\widehat{\Gamma}_i \ \ \mbox{are defined by (\ref{ene23})})
$$
is a self-adjoint extension ${A}_{{\mathbf{B}}}$ of the nondensely
defined operator $A_M=-d^2/dx^2+I$, \
$\mathcal{D}(A_M)=\{u(x)\in{W}_2^{p+2}(\mathbb{R}) \ | \
u(0)=\ldots=u^{(p+1)}(0)=0\}$ acting in $W^p_2(\mathbb{R})$. The
operator $A_B$ can be interpreted as a one-dimensional
Schr\"{o}dinger operator with point interaction in the Sobolev space
$W_2^p(\mathbb{R})$ \cite{Nizh}.

\subsection{Realization of self-adjoint extensions of $A_M$ in $\mathcal{H}_p$
by additive perturbations.}

In mathematical physics, the self-adjoint extensions
$A_{\mathbf{B},\mathbf{R}}$ of $A_N$ described  in Theorem
\ref{ttt12} appear naturally as self-adjoint realizations of the
additive purely singular perturbations (\ref{ne3}) in $\mathcal{H}$.
Our aim is to give a similar interpretation for self-adjoint
extensions $A_{{B}}$ of $A_M$ defined by (\ref{tat77}) in the space
$\mathcal{H}_p$. In what follows, without loss of generality, we
assume that an auxiliary Hilbert space $\mathfrak{N}$ in
$(\oplus\mathfrak{N}^{p/2+1}, \widehat{\Gamma}_0,
\widehat{\Gamma}_1)$ coincides with $\mathbb{C}^n$ (here
$n=\dim{\mathfrak{N}}$). So,
$\oplus\mathfrak{N}^{p/2+1}=\mathbb{C}^{n(p/2+1)}$. In this case the
operator $B$ in (\ref{tat77}) is given by an Hermitian matrix
$\mathbf{B}$ of the order $n(p/2+1)$.

It follows from the relation $\ker\Gamma_1=\mathcal{D}(A)$ and
equalities (\ref{ee5}), (\ref{tat17}) that
$\ker\widehat{\Gamma}_1=\mathcal{D}(A^{p+1})$. Hence, the
restriction $\widehat{\Gamma}_1\upharpoonright_{M}$ determines a
one-to-one correspondence between $M$ and $\mathbb{C}^{n(p/2+1)}$.
Thus $(\widehat{\Gamma}_1\upharpoonright_{M})^{-1}$ exists and
$(\widehat{\Gamma}_1\upharpoonright_{M})^{-1}$ maps
$\mathbb{C}^{n(p/2+1)}$ onto $M$.

 Putting
$\Psi{d}:=-A^{p+1}(\widehat{\Gamma}_1\upharpoonright_{M})^{-1}d$,
where $d\in\mathbb{C}^{n(p/2+1)}$, we determine an injective linear
mapping of $\mathbb{C}^{n(p/2+1)}$ to $\mathcal{H}_{-p-2}$ such that
$\mathcal{R}(\Psi)\cap\mathcal{H}=\{0\}$.

Let us determine its adjoint
$\Psi^*:\mathcal{H}_{p+2}\to\mathbb{C}^{n(p/2+1)}$ by the formula
\begin{equation}\label{tyt21}
 <u, \Psi{d}>=(\Psi^*u,d)_{\mathbb{C}^{n(p/2+1)}}, \ \ \ \
\forall{u}\in\mathcal{H}_{p+2}=\mathcal{D}(A^{p/2+1}), \ \
\forall{d}\in\mathbb{C}^{n(p/2+1)}.
\end{equation}

To describe $\Psi^*$ we set $\psi_j=\Psi{e_j}$, where
$\{e_j\}_1^{n(p/2+1)}$ is the canonical basis of
$\mathbb{C}^{n(p/2+1)}$. Setting $f=u\in\mathcal{D}(A^{p+1})$ and
$g=A^{-p-1}\psi_j=A^{-p-1}\Psi{e_j}=$ \linebreak[4]
$-(\widehat{\Gamma}_1\upharpoonright_{M})^{-1}e_j$ in the Green
identity (\ref{tyt9}), using (\ref{e2233}), and recalling that
$\ker\widehat{\Gamma}_1=\mathcal{D}(A^{p+1})$, we get
$$
<u,\psi_j>=(A^{p+1}u, A^{-p-1}\psi_j)=
 -(\widehat{\Gamma}_0u,\widehat{\Gamma}_1g)_{\mathbb{C}^{n(p/2+1)}}=
(\widehat{\Gamma}_0u,e_j)_{\mathbb{C}^{n(p/2+1)}}.
$$
The latter relation and (\ref{tyt21}) imply that
\begin{equation}\label{tyt166}
\Psi^*{u}=\left(\begin{array}{c}
 <u, \psi_1> \\
 \vdots \\
<u, \psi_{n(p/2+1)}>
\end{array}\right)=\widehat{\Gamma}_0u
\end{equation}
for `smooth' vectors
${u}\in\mathcal{D}(A^{p+1})=\mathcal{H}_{2p+2}.$ The continuation of
$\Psi^*$ onto $\mathcal{D}(A^{p/2+1})=\mathcal{H}_{p+2}$ is obtained
by the closure.

Let us consider the formal expression
 \begin{equation}\label{tyt15}
 A_p +\sum_{i,j=1}^{n(p/2+1)}{b}_{ij}<\cdot,\psi_j>\psi_i=A_p+\Psi\mathbf{B}\Psi^{*},
 \end{equation}
where $\mathbf{B}=(b_{ij})_{ij}^{n(p/2+1)}$ is an Hermitian matrix
of the order $n(p/2+1)$ and
$A_p=A\upharpoonright_{\mathcal{D}(A^{p/2+1})}$ is a self-adjoint
operator in $\mathcal{H}_p$.

In general, the singular elements $\psi_j$ belong to
$\mathcal{H}_{-p-2}$ and hence, they are well defined on
$u\in\mathcal{H}_{p+2}$ . For this reason it is natural to consider
the `potential' $V=\Psi\mathbf{B}\Psi^{*}$ in (\ref{tyt15}) as a
singular perturbation of the `free' operator $A_p$  in
$\mathcal{H}_p$ and, reasoning by analogy with Subsection 2.2.1, to
give a meaning of the formal expression (\ref{tyt15}) as a
self-adjoint operator extension $\widetilde{A}$ of the symmetric
operator (cf. (\ref{e7}))
$$
A_{\mathrm{sym}}:=A_p\upharpoonright_{{\mathcal{D}}(A_{\mathrm{sym}})},
\ \ \ {\mathcal{D}}(A_{\mathrm{sym}})=\{u\in\mathcal{D}(A^{p/2+1}) \
| \ \Psi^{*}u=0 \}
$$
acting in $\mathcal{H}_{p}$.

It follows from (\ref{tyt20}) and (\ref{tyt21}) that
$A_{\mathrm{sym}}=A_M$. So, in contrast to the operator
$A_{\mathrm{sym}}=A_N$ defined by (\ref{e7}), the operator
$A_{\mathrm{sym}}=A_M$ {\it is non-densely defined}. Therefore, a
modification of the Albeverio-Kurasov approach (see Subsection
2.2.1) is required to describe self-adjoint extensions of $A_M$ by
additive mixed singular perturbation (\ref{tyt15}).

First of all we restrict (\ref{tyt15}) to the set
$\mathcal{D}(A^{p+1})$ and define the action of (\ref{tyt15}) on
vectors from the domain of definition
$\mathcal{D}(A_M^{(*)})=\mathcal{D}(A^{p+1})\dot{+}M$ of the
quasi-adjoint operator $A_M^{(*)}$ (in other words, we construct a
regularization $A^+_p+\Psi\mathbf{B}\Psi^{*}_{\mathbf{R}}$ of
(\ref{tyt15}) defined on $\mathcal{D}(A^{p+1})\dot{+}M$).

Relation (\ref{tyt166}) means that the extension
$\Psi^*_{\mathbf{R}}$ can naturally be defined by the boundary
operator $\widehat{\Gamma}_0$. Namely,
\begin{equation}\label{tyt10}
\Psi^*_{\mathbf{R}}{f}=\left(\begin{array}{c}
<f, \psi_1^{\mathrm{ex}}>  \\
 \vdots \\
<f, \psi_{n(p/2+1)}^{\mathrm{ex}}>
\end{array}\right):=\widehat{\Gamma}_0f, \qquad \forall{f}\in\mathcal{D}(A^{p+1})\dot{+}M.
\end{equation}

The extension $A^{+}_p$ of $A_p$ can be defined by analogy with
(\ref{tat101}). Precisely, we only need to indicate the action of
$A^+_p$ on $M$.  Assuming that $A^{+}_p\upharpoonright_{M}$ acts as
the isometric mapping $A^{p+1}$ in $A$-scale (see Subsection 2.2),
we get
\begin{equation}\label{tat111}
A^+_pf=A_pu+A^{p+1}\mathfrak{m}=A_M^{(*)}f+A^{p+1}\mathfrak{m}, \ \
\ \forall{f}=u+\mathfrak{m}\in\mathcal{D}(A_M^{(*)}).
\end{equation}
After such a preparation work, the operator realization
$\widetilde{A}$ of (\ref{tyt15}) in $\mathcal{H}_p$  is determined
by the formula (cf. (\ref{lesia40}))
\begin{equation}\label{sasha}
\widetilde{A}=[A^+_p+\Psi\mathbf{B}\Psi^{*}_{\mathbf{R}}]\upharpoonright_{\mathcal{D}(\widetilde{A})},
\quad \mathcal{D}(\widetilde{A})=\{f\in\mathcal{D}(A^{p+1})\dot{+}M
\ | \ A^+_pf+\Psi\mathbf{B}\Psi^{*}_{\mathbf{R}}f\in\mathcal{H}_p\}.
\end{equation}

\begin{ttt}\label{t99}
Let ${\mathbf{B}}$ be an Hermitian matrix of the order $n(p/2+1)$.
Then the operator $\widetilde{A}$ is essentially self-adjoint in
$\mathcal{H}_p$ and it can be also defined by the formula
\begin{equation}\label{sasha1}
{A}_{\mathbf{B}}'=A_{M}^{(*)}\upharpoonright_{\mathcal{D}({A}_{\mathbf{B}}')},
\ \ \
\mathcal{D}({A}_{{\mathbf{B}}}')=\{f\in\mathcal{D}(A^{p+1})\dot{+}M
\ | \ \mathbf{B}\widehat{\Gamma}_0f= \widehat{\Gamma}_1f\}.
\end{equation}
\end{ttt}

{\it Proof.} By the definition of $\Psi$,
$A^{p+1}\mathfrak{m}=-\Psi\widehat{\Gamma}_1\mathfrak{m}=-\Psi\widehat{\Gamma}_1f$
for any $\mathfrak{m}\in{M}$ and $f=u+\mathfrak{m}$ \
$(u\in\mathcal{D}(A^{p+1}))$. The obtained expression,
(\ref{tyt10}), and (\ref{tat111}) yield
\begin{equation}\label{sasha4}
[A^+_p+\Psi\mathbf{B}\Psi^{*}_{\mathbf{R}}]f=A_M^{(*)}f+\Psi[\mathbf{B}\widehat{\Gamma}_0-\widehat{\Gamma}_1]f
\ \ \ \ (\forall{f}\in\mathcal{D}(A^{p+1})\dot{+}M).
\end{equation}
The latter equality means
$f\in\mathcal{D}(\widetilde{A})\iff{\mathbf{B}\widehat{\Gamma}_0f=\widehat{\Gamma}_1f}$
(since $\mathcal{R}(\Psi)\cap\mathcal{H}=\{0\}$ and hence,
$\mathcal{R}(\Psi)\cap\mathcal{H}_p=\{0\}$). Combining this fact
with (\ref{sasha}) -- (\ref{sasha4}) we conclude that
$\widetilde{A}$ coincides with ${A}_{\mathbf{B}}'$. The property of
the operator $\widetilde{A}$ to be essentially self-adjoint follows
from Theorem \ref{p25}.

\section{Acknowledgments} The
second (S.K) and third (L.N.) authors thank DFG for the financial
support of the projects 436 UKR 113/88/0-1 and 436 UKR 113/79,
respectively, and the Institute f\"{u}r Angewandte Mathematik der
Universit\"{a}t Bonn for the warm hospitality.
 
 \end{document}